\documentclass[]{jfm}

\usepackage{graphicx}
\usepackage{epstopdf,epsfig}

\usepackage{newtxtext}
\usepackage{newtxmath}
\usepackage{natbib}

\usepackage{amsmath}
\usepackage{amssymb,amsbsy,bm}
\usepackage{comment} 
\usepackage[dvipsnames]{xcolor}

\usepackage[percent]{overpic} 
\usepackage{tikz}
\usetikzlibrary{arrows.meta}

\usepackage{subfiles}
\usepackage{subfig}

\usepackage{placeins}
\usepackage{mathtools,xparse}

\usepackage{hyperref}
\hypersetup{
    colorlinks = true,
    urlcolor   = blue,
    citecolor  = blue,
    linkcolor=red,
    citecolor=blue,
    filecolor=cyan,
}

\usepackage{cleveref}
\usepackage[normalem]{ulem}

\crefformat{section}{\S#2#1#3} 
\crefformat{subsection}{\S#2#1#3}
\crefformat{subsubsection}{\S#2#1#3}

\usepackage{array}

\newcommand{\RomanNumeralCaps}[1]
\linenumbers

\newcommand{\St}{\textit{St}}
\newcommand{\Fr}{\textit{Fr}}
\newcommand{\Sv}{\textit{Sv}}
\newcommand{\of}[1]{\left(#1\right)}

\newcommand{\vect}[1]{\bm{#1}}

\title{How two-way coupling modifies the multiscale preferential sweeping mechanism}

\author{Josin Tom \aff{1},
  Maurizio Carbone \aff{2},
  Andrew D. Bragg \aff{1}
  \corresp{\email{andrew.bragg@duke.edu}}
}
\affiliation{
\aff{1}
Department of Civil and Environmental Engineering, Duke University, Durham, NC 27708, USA
\aff{2}
Max Planck Institute for Dynamics and Self-Organization, Am Fa{\ss}berg 17, 37077 Göttingen, Germany
}

\begin{document}
\maketitle

\begin{abstract}

For one-way coupled (1WC) flows, Tom \& Bragg (J. Fluid Mech., 871, pp. 244-270, 2019) advanced the analysis of Maxey (J. Fluid Mech., 174, pp. 441–465, 1987), which applied to weakly inertial particles, to particles of arbitrary inertia, and the new theoretical result revealed the role that different scales play in the preferential sweeping mechanism that leads to enhanced particle settling in turbulent flows. Monchaux \& Dejoan (Phys. Rev. Fluids, 2, 104302, 2017) showed using direct numerical simulations (DNS) that while for low particle loading the effect of two-way coupling (2WC) on the global flow statistics is weak, 2WC enables the particles to drag the fluid in their vicinity down with them, significantly enhancing their settling, and they argued that 2WC suppresses the preferential sweeping mechanism. We explore this further by considering the impact of 2WC on the contribution made by eddies of different sizes on the particle settling. In agreement with Monchaux \& Dejoan, we show that even for low loading, 2WC strongly enhances particle settling, and we show how 2WC modifies the contribution from different flow scales. However, contrary to their study, we show that preferential sweeping remains important in 2WC flows. In particular, for both 1WC and 2WC flows, the settling enhancement due to turbulence is dominated by contributions from particles in straining regions of the flow, but for the 2WC case, the particles in these regions also drag the fluid down with them, leading to an enhancement of their settling compared to the 1WC case.

\end{abstract}


{\bf MSC Codes }  {\it(Optional)} Please enter your MSC Codes here

\section{Introduction}\label{sec:Intro}
Inertial particles settling in turbulent flows are important in many environmental and biological problems, ranging from cloud microphysics \citep{prupp97,grabowski13}, dispersion of pollutants in the atmosphere \citep{dhariwal19,Toscano2021} to aerosol deposition in human lungs \citep{chen16,ou2020}. Understanding the physical mechanisms responsible for the modification of the particle settling velocity due to turbulence and its dependence on the system parameters is an important and ongoing research topic. Even for the simplest case of small, heavy, spherical inertial particles settling in homogeneous turbulent flows, the problem is very challenging to understand, and there remain many unanswered questions.

One key feature that has frequently been observed is that inertial particles settling in turbulence fall faster on average than they would in a quiescent fluid \citep{maxey87,wang93,aliseda02,good14,ireland16b,petersen19,momenifar19a,momenifar19b}. It has also been suggested that turbulence could lead to a reduction in the particle settling velocity, an effect referred to as `loitering' \citep{nielsen93}. Although there is some evidence of loitering occuring in certain parameter regimes, e.g.~see \cite{good14}, this remains controversial, and the majority of the evidence points to a settling enhancement, rather than reduction. \cite{good14} suggested, by comparing direct numerical simulations (DNS) of small, heavy inertial particles using linear and nonlinear drag, that loitering can only occur if the particles experience significant nonlinear drag forces. However, a subsequent DNS study by \cite{rosa16} called that conclusion into question, finding that linear and nonlinear drag models gave virtually identical results showing only enhanced settling due to turbulence. \cite{rosa16} suggested that a possible explanation for the discrepancy is that the DNS of \cite{good14} were not run for sufficiently long times for the particle settling velocities to statistically converge. This is a point we will return to later.

The particle inertia can be quantified by the Kolmogorov scale Stokes number, $St$, that is defined as the ratio between the particle response time $\tau_p$ and the Kolmogorov time $\tau_\eta$.
The settling number $\Sv\equiv\tau_p g /u_\eta$ quantifies the particle settling velocity in a quiescent fluid (i.e.~the Stokes settling velocity $\tau_p g$) relative to the Kolmogorov velocity $u_\eta$.
One may also use the Froude number $\Fr\equiv a_\eta/g$ that compares the Kolmogorov acceleration to that of gravity.
The two numbers, $\St$ and $\Sv=\St/\Fr$, may be used to fully characterize an idealized one-way-coupled (1WC) system, in which the carrier fluid drives the motion of the particles, while the force from the particles on the fluid is assumed negligible. Such an idealized system is thought to be a good approximation when the particle mass loading is sufficiently small. A two-way-coupled (2WC) system is characterized by additional parameters such as the particle volume fraction $\Phi_v\equiv (\pi/6) N_p (d_p/L)^3$ (for spherical particles in a cubic flow domain), where $d_p$ is the particle diameter, $N_p$ is the number of particles in the flow, $L$ is the length of the flow domain, and the mass fraction, $\Phi_m\equiv(\rho_p /\rho_f) \Phi_v$, where $\rho_p /\rho_f$ is the ratio of the particle density to the fluid density. We are concerned with dilute suspensions, i.e. $\Phi_v<10^{-3}$, of particles that are small, meaning $d_p/\eta \ll 1$ (where $\eta$ is the Kolmogorov length scale), and heavy, meaning $\rho_p/\rho_f\gg 1$, which is appropriate for diverse physical systems, from droplets dynamics in atmospheric clouds \citep{prupp97} to the transport of pharmaceutical nasal sprays \citep{kolanjiyil2021}.

In the context of 1WC systems, a key study that provided new insights into the physical mechanism by which turbulence modifies the settling velocities of inertial particles was that of \cite{maxey87}, who showed that in the presence of gravity, particles tend to fall around the downward moving side of vortices in the flow. As a result of this, they fall faster on average in a turbulent flow than they would in a quiescent flow, an effect later referred to as the ``preferential sweeping'' mechanism \citep{wang93}. The preferential sweeping mechanism arises due to two effects in the system. First, \cite{maxey87} argued that inertial particles are centrifuged out from vortices in the flow, causing them to preferentially sample strain-dominated regions. This was subsequently observed and confirmed using DNS by \cite{squires90}. Second, \cite{maxey87} argued that while the centrifuge effect causes the particles to move around the vortices in the flow, gravity causes them to be preferentially swept around the downward side of the vortices (because that is the `path of least resistance’ as pointed out in \citet{tom19}), such that they preferentially follow regions of the flow where the fluid velocity points in the direction of gravity. This preferential sweeping effect was observed and confirmed using DNS in \cite{wang93}. Several subsequent numerical and experimental studies provided confirmation of settling enhancement due to the preferential sweeping effect \citep{bec14b,ireland16b,rosa16,petersen19,li21b}. In the original study of \cite{maxey87}, the preferential sweeping mechanism was described theoretically for particles with weak inertia, i.e.~$\St\ll 1$. The theory was recently extended to finite Stokes numbers by \cite{tom19}, revealing the role played in the preferential sweeping mechanism by vortices of different sizes and the $St$ and $Fr$ dependence of the `multiscale preferential sweeping' mechanism. It should also be noted that most of these studies focused on homogeneous turbulence (which is also the case of interest for our study), but recently the role of preferential sweeping has been extended to the case of wall-bounded turbulence \citep{bragg21a,bragg21b}. These studies have shown that throughout much of the boundary layer, the preferential sweeping mechanism continues to play an important role. However, very close to the wall (e.g.~in the buffer and viscous sublayers) where the turbulence inhomogeneity is strongest, the turbophoretic drift mechanism becomes the dominant mechanism responsible for enhancing the particle settling velocity. 

As noted previously, except in restricted regimes, two-way coupling (2WC) can also be important for how the inertial particles settle. It is commonly argued that the effects of 2WC are only important when the volume loading exceeds a certain threshold, e.g.~when $\Phi_v\geq 10^{-6}$ \footnote[1]{Strictly speaking, it is $\Phi_m$ that determines whether the effects of 2WC are important, not $\Phi_v$. However, since they are proportional for a given $\rho_p/\rho_f$, then often the influence of the particles is discussed in terms of $\Phi_v$.}. For example, in \cite{bosse06} the fluid energy spectrums for 1WC and 2WC flows are almost identical for $\Phi_v = 10^{-5}$,  but become noticeably different for $\Phi_v = 10^{-4}$. However, \cite{monchaux17} showed using DNS that even when $\Phi_v$ was small enough such that the globally averaged fluid statistics are almost the same for the 1WC and 2WC cases, the average particle settling velocities were nevertheless very strongly modified due to 2WC. Their explanation was that even though $\Phi_v$ was too small for the inertial particles to significantly affect the global fluid statistics, 2WC can nevertheless strongly modify the flow statistics in the vicinity of the particles, which in turn affects the particle settling due to modifications of the drag force acting on the particles.  Therefore, 2WC may be important for particle processes such as settling, in regimes where it was previously thought to be unimportant.

Concerning the mechanism responsible for enhanced particle settling in turbulent flows with 2WC, numerical simulations with 2WC \citep{bosse06,monchaux17} and laboratory experiments \citep{good14,petersen19} have shown that clusters of particles tend to fall faster than particles that are not part of a cluster. In particular, particles fall faster than average when the local particle number density is larger than $\approx 3$ times the average number density. This is possibly consistent with the preferential sweeping effect since particle clusters would be expected to form in the same strain-dominated regions where the particles preferentially sample the downward moving fluid. However, it could also be the case that the settling is faster in these regions because 2WC allows the particle clusters to drag the surrounding fluid down with then, decreasing the drag resistance to their falling motion \citep{monchaux17}. Indeed, by considering the flow topology sampled by the settling particles, \citet{monchaux17} argued that as $\Phi_m$ is increased, the preferential sweeping effect becomes less important, and the dragging of the fluid by the particle becomes the main mechanism associated with the enhanced particle settling velocities. It is also worth mentioning that exciting new field experiments have provided evidence for settling enhancement due to turbulence in diverse naturally occurring flows \citep{nemes17,li21a}. Moreover, \cite{li21b} provide direct evidence that this enhancement is, in fact, associated with the preferential sweeping mechanism. Since these field experiments could be impacted by 2WC, the results may suggest that the preferential sweeping mechanism still applies in that regime. 

While the aforementioned studies point to an important role from 2WC in determining the particle settling speeds, a number of interesting and important issues remain to be solved. In \cite{tom19}, we used theory and DNS to reveal the truly multiscale character of the preferential sweeping mechanism, and the way in which a restricted range of turbulent eddies contributes to the preferential sweeping of an inertial particle, with the relevant range depending upon $\St$ and $\Sv$. A natural and important question is how 2WC modifies this multiscale physical picture. It is precisely this question that we explore in the present paper, by extending the analysis of \cite{tom19} to a system with 2WC. One important finding of our study is that it yields an alternative interpretation to that presented in \cite{monchaux17} regarding the role of preferential sweeping in the presence of 2WC.

The outline of the rest of the paper is as follows. In \S\ref{sec:Theory} we summarize the theory of \cite{tom19} and discuss its extension to 2WC flows. In \S\ref{sec:DNS_Details} we summarize the details of our DNS and the method used to capture 2WC. In \S\ref{sec:DNS_Results} we present and discuss the DNS results and explore various quantities in order to understand the role that 2WC plays in the enhancement of particle settling velocities due to turbulence and its impact on the preferential sweeping mechanism. Finally, in \S\ref{sec:Conclusions} we draw conclusions to our work and mention important steps for future work.

%
%
\section{Theory}\label{sec:Theory}
\subsection{Background}\label{Background}
We consider the settling of small ($d_p/\eta \ll 1$, where $d_p$ is the particle diameter and $\eta$ is the Kolmogorov length scale), heavy ($\rho_p/\rho_f \gg 1$, where $\rho_p$ is particle density and $\rho_f$ is fluid density), spherical, mono-disperse inertial particles in a statistically stationary, homogeneous (and initially isotropic) turbulent flow. In the regime of a linear drag force, the particle equation of motion reduces to \citep[see][]{maxey83} 
\begin{align}
\ddot{\bm{x}}^p(t)\equiv\dot{\bm{v}}^p(t)=\frac{1}{St\tau_\eta}\Big(\bm{u}(\bm{x}^p(t),t)-\bm{v}^p(t)\Big)+\bm{g},\label{eqn_eom}
\end{align}
where $\bm{x}^p(t),\bm{v}^p(t)$ are the particle position and velocity vectors, respectively, $\bm{u}(\bm{x}^p(t),t)$ is the undisturbed fluid velocity at the particle position, and $\bm{g}$ is the gravitational acceleration vector. Even though idealized, this simplified equation of motion is widely used for describing particle motion in atmospheric flows (e.g. \cite{grabowski13}), and even for this equation of motion, many unanswered questions remain concerning the settling of particles in turbulence. Moreover, our DNS will be confined to $St\leq 2$ and $Sv\leq 2$ for which we expect that the linear drag approximation will be reasonable.

In a 2WC system, a particle exerts a force on the fluid at its location equal to 
\begin{align}
\bm{f}^p(t) = -\frac{m_p}{St\tau_\eta}\Big(\bm{u}(\bm{x}^p(t),t)-\bm{v}^p(t)\Big),\label{eqn_f_feedbk}
\end{align}
which is equal and opposite to the drag force exerted by the fluid on the particles that is described by \eqref{eqn_eom}, and $m_p=(4/3)\pi (d_p/2)^3\rho_p $ is the mass of the spherical particle. The addition of the feedback force, $\bm{f}^p(t)$, breaks the isotropic symmetry of an otherwise isotropic turbulent flow, but the fluid is still statistically homogeneous since the gravitational body force is independent of position. 

Statistical stationarity of the system implies that, in the vertical direction defined by $\bm{e}_z$, the ensemble average of \eqref{eqn_eom} is (with $\bm{g}=-g\bm{e}_z$)
\begin{align}
\langle{v}_z^p(t)\rangle=\langle{u}_z(\bm{x}^p(t),t)\rangle-St\tau_\eta{g}.\label{eqn_SV}
\end{align}
Note that we used $\bm{g}=g\bm{e}_z$ in \citet{tom19}, but have decided to switch to this more standard choice of coordinate system in this paper. Equation \eqref{eqn_SV} shows that the average particle velocity may differ from the Stokes settling velocity $-St\tau_\eta{g}$ only if $\langle{u}_z(\bm{x}^p(t),t)\rangle\neq{0}$ (we consider flows with Eulerian average $\langle u_z(\bm{x},t)\rangle=0$, such that $\langle{u}_z(\bm{x}^p(t),t)\rangle\neq{0}$ can only arise due to non-trivial effects in the flow). Several numerical studies for both 1WC \citep{wang93,bec14b,ireland16b,rosa16} and 2WC systems \citep{bosse06,dejoan11,monchaux17,rosa21} have indeed shown that $\langle {v}_z^p(t)\rangle\neq -St\tau_\eta{g}$, implying $\langle{u}_z(\bm{x}^p(t),t)\rangle\neq{0}$. Experimental studies in the laboratory \citep{aliseda02,yang05,good14,petersen19} and more recently, settling velocity measurements in the atmosphere \citep{nemes17,li21a,li21b}, both of which inherently have 2WC effects, have also observed an enhancement of settling velocity compared to the Stokes settling velocity. Hence, there is a need to explore the physical mechanism(s) responsible for generating $\langle{u}_z(\bm{x}^p(t),t)\rangle<{0}$ and whether there are different mechanisms at play for the 1WC and 2WC systems. 

\subsection{Brief summary of the theoretical framework}\label{Theory_Frame}

For the 1WC system, \citet{maxey87} developed a theoretical framework to explain how $\langle{u}_z(\bm{x}^p(t),t)\rangle<{0}$ can occur even in flows where the Eulerian flow average is $\langle{u}_z(\bm{x},t)\rangle={0}$. Essentially, the explanation is that particles with inertia do not uniformly sample the underlying fluid velocity field, and that gravity leads to a bias for inertial particles to accumulate in regions of the flow where $\bm{e}_z\bm{\cdot u}<0$. However, the analysis of \citet{maxey87} was restricted to $St\ll1$. In a recent paper \citep{tom19}, we developed a theoretical framework for considering the behavior of $\langle{u}_z(\bm{x}^p(t),t)\rangle$ for arbitrary $St$, and for revealing the scales of motion that contribute to $\langle{u}_z(\bm{x}^p(t),t)\rangle$ being finite and negative. Here, we briefly summarize the steps of the theoretical analysis and comment on how the analytical result can be used to understand how 2WC might modify the physical mechanisms governing the particle settling velocity.

The key ideas used for developing the theoretical framework to analyze $\langle{u}_z(\bm{x}^p(t),t)\rangle$ for arbitrary $St$ are averaging decompositions and Probability Density Function (PDF) methods. The only assumptions made in our derivation are that the system is statistically stationary and homogeneous, and that the particle dynamics are governed by \eqref{eqn_eom}. The basic steps in the derivation are as follows (see \cite{tom19} for more details). For the system under consideration, an ensemble average $\langle\cdot\rangle$ over all possible states of the system corresponds to averaging over all realizations of $\bm{u}$, as well as over all initial particle positions $\bm{x}^p(0)=\bm{x}_0$ and velocities $\bm{v}^p(0)=\bm{v}_0$. In view of this, we introduce the averaging decomposition $\langle\cdot\rangle=\langle\langle\cdot\rangle^{\bm{x}_0,\bm{v}_0}_{\bm{u}}\rangle^{\bm{u}}$ \citep{bragg12b}, where $\langle\cdot\rangle^{\bm{x}_0,\bm{v}_0}_{\bm{u}}$ denotes an average over all initial particle positions $\bm{x}^p(0)=\bm{x}_0$ and velocities $\bm{v}^p(0)=\bm{v}_0$ for a given realization of the fluid velocity field $\bm{u}$, and $\langle\cdot\rangle^{\bm{u}}$ denotes an average over all realizations of $\bm{u}$. Using this operator, we construct a generalized particle velocity field $\bm{\mathcal{V}}(\bm{x},t)\equiv  \langle\bm{v}^p(t)\rangle^{\bm{x}_0,\bm{v}_0}_{\bm{u}}$ that is well-defined for all $St$, including the regime where caustics occur \citep{wilkinson05} because it is formed by averaging over trajectories satisfying $\bm{x}^p(t)=\bm{x}$ (i.e. it does not assume uniqueness of these trajectories, a scenario that only applies in the limit $St\to 0$ that was considered in \cite{maxey87}). With this velocity field we then construct the result
\begin{align}
\langle{u}_z(\bm{x}^p(t),t)\rangle=\Bigg\langle{u}_z(\bm{x},t)\exp\Bigg( -\int_0^t \bm{\nabla\cdot}\bm{\mathcal{V}}(\bm{\mathcal{X}}(s\vert\bm{x},t),s)\,ds \Bigg)\Bigg\rangle^{\bm{u}},\label{GR2}
\end{align}
where the notation $s\vert\bm{x},t$ denotes that the variable is measured at time $s$ along a trajectory satisfying $\bm{\mathcal{X}}(t)=\bm{x}$, and where $\partial_s\bm{\mathcal{X}}\equiv\bm{\mathcal{V}}(\bm{\mathcal{X}}(s),s)$.

The result was further developed to gain insight into the multiscale nature of the problem by introducing coarse-graining decompositions $u_z(\bm{x},t)=\widetilde{u}_z(\bm{x},t)+u'_z(\bm{x},t)$ and $\bm{\mathcal{V}}=\widetilde{\bm{\mathcal{V}}}+\bm{\mathcal{V}}'$, where $\widetilde{u}_z$ and $\widetilde{\bm{\mathcal{V}}}$ denote the fields $u_z$ and ${\bm{\mathcal{V}}}$ coarse-grained on the length scale $\ell_c$, while $u'_z(\bm{x},t)\equiv u_z(\bm{x},t)-\widetilde{u}_z(\bm{x},t)$ and $\bm{\mathcal{V}}'\equiv\bm{\mathcal{V}}-\widetilde{\bm{\mathcal{V}}}$ are the `sub-grid' fields. Next, we choose $\ell_c$ to be a function of $\St$, i.e. $\ell_c(\St)$, which is defined such that the scale-dependent Stokes number $St_{\ell}\equiv\tau_p/\tau_{\ell}$ (where $\tau_{\ell}$ is the eddy-turnover timescale at scale $\ell$) is $\lll1$ for $\ell>\ell_c(St)$. Physically, this means that $\ell_c(St)$ is defined to be the scale above which the particle behaves as if it were a fluid particle. Using this choice for the filtering length, we finally obtain

\begin{align}
\begin{split}
\langle{u}_z(\bm{x}^p(t),t)\rangle \approx \Bigg\langle{u}'_z(\bm{x},t)\exp\Bigg( -\int_0^t\bm{\nabla\cdot}\bm{\mathcal{V}}'(\bm{\mathcal{X}}(s\vert\bm{x},t),s)\,ds \Bigg)\Bigg\rangle^{\bm{u}}.
\end{split}
\label{Alt}
\end{align}

Since the RHS of this result only contains the sub-grid fields, it shows that the turbulent flow scales that contribute to the enhanced particle settling speeds are those with size $<\ell_c$, while scales $>\ell_c$ make a vanishingly small contribution. The physical mechanism embedded in \eqref{Alt} is a multiscale version of the original preferential sweeping mechanism described by \cite{maxey87} and \cite{wang93}. In particular, according to \eqref{Alt}, $\langle{u}_z(\bm{x}^p(t),t)\rangle<0$ occurs because the inertial particles are preferentially swept by eddies of size $\ell <\ell_c(St)$.

\subsection{Applicability of theoretical analysis the for 2WC case}\label{Theory_2WC_Appl}

Although the result in \eqref{Alt} was derived for a 1WC system, its assumptions are such that it also applies in the 2WC regime. The only difference is one of interpretation, since in the 2WC case, all of the dynamical variables contained within \eqref{Alt} are implicitly affected by the force from the particle on the fluid. In view of this, we now consider how 2WC might affect the different quantities in \eqref{Alt} in order to understand the impact of 2WC on particle settling in turbulence.

The result in \eqref{Alt} shows that the settling modification due to turbulence associated with $\langle{u}_z(\bm{x}^p(t),t)\rangle<0$ is determined by two things, namely the strength of the fluctuations of the sub-grid field ${u}'_z(\bm{x},t)$, and the intensity of the particle clustering (associated with the exponential term involving the compressibility $\bm{\nabla\cdot}\bm{\mathcal{V}}'$). It is to be noted, however, that as emphasized in \cite{tom19}, it is not merely the intensity of the clustering that matters but also whether it is correlated with ${u}'_z(\bm{x},t)$. For example, in the 1WC case, unless the clustering is correlated with ${u}'_z(\bm{x},t)$ then the RHS of \eqref{Alt} is zero because $\langle {u}'_z(\bm{x},t)\rangle^{\bm{u}}=0$. For this reason, it is better to think of the exponential integral term in \eqref{Alt} as describing the process of preferential concentration rather than clustering, where preferential concentration describes how particles cluster preferentially in certain regions of the flow (see \cite{bragg14d} for a discussion on the subtle differences between these phenomena).

In the 1WC regime, the particles do not modify ${u}'_z(\bm{x},t)$. In the 2WC regime, the particles will modify ${u}'_z(\bm{x},t)$, and previous results for non-settling particles have revealed a `pivoting effect' wherein the inertial particles were observed to increase the energy content of turbulent flow scales of size less than some threshold, and decrease the energy content of flow scales larger than this \citep{elghobashi93,sundaram99b,bosse06}. The result in \eqref{Alt} shows that flow scales of size $\ell<\ell_c(St)$ contribute to the particle settling. The flows scales in the range  $\ell<\ell_c(St)$ that dominate the settling will determine whether the pivoting mechanism tends to decrease or increase ${u}'_z(\bm{x},t)$. In the context of gravitational settling, \citet{monchaux17} argued that the settling particles can drag the fluid in their vicinity with them, an effect referred to as the `fluid-drag mechanism'. This fluid drag mechanism could lead to an increase in the negative values of ${u}'_z(\bm{x},t)$ responsible for generating $\langle{u}_z(\bm{x}^p(t),t)\rangle<0$. In such a case, 2WC could enhance the particle settling speeds.

2WC can also, however, modify the preferential concentration of the particles associated with the contribution $\exp( -\int_0^t\bm{\nabla\cdot}\bm{\mathcal{V}}'(\bm{\mathcal{X}}(s\vert\bm{x},t),s)\,ds)$ in \eqref{Alt}. It is more difficult to predict theoretically how 2WC might affect this contribution to \eqref{Alt}, however this is something that we can explore using DNS. It is important to note, however, that in a 2WC system, enhanced settling could occur even in the absence of preferential concentration. For example, even if $\exp( -\int_0^t\bm{\nabla\cdot}\bm{\mathcal{V}}'(\bm{\mathcal{X}}(s\vert\bm{x},t),s)\,ds)= 1$ for all $\bm{x}^p(t)=\bm{x}$, then we would have
\begin{align}
\begin{split}
\langle{u}_z(\bm{x}^p(t),t)\rangle \approx \Big\langle{u}'_z(\bm{y},t)\Big\rangle^{\bm{u}},
\end{split}
\label{Altb}
\end{align}
where $\bm{y}$ corresponds to the homogeneous distribution of points in space where the particles are located. Although in a 1WC system $\langle{u}'_z(\bm{y},t)\rangle^{\bm{u}}=0$, in a 2WC system we could have  $\langle{u}'_z(\bm{y},t)\rangle^{\bm{u}}<0$ and hence $\langle{u}_z(\bm{x}^p(t),t)\rangle <0$ due to the settling particles dragging the fluid down with them. In such a scenario the preferential sweeping mechanism would be playing no role. Measuring preferential concentration in the flow can therefore be used to indirectly determine whether $\langle{u}_z(\bm{x}^p(t),t)\rangle<0$ in a 2WC system is due to preferential sweeping or not. 

In our previous work \citep{tom19}, we studied the contributions of different flow scales to the settling velocity enhancement in a 1WC system. In this work, we aim to extend this analysis to the 2WC regime. First, we will look at the effect of 2WC on the overall settling velocity enhancement, i.e. the contribution of all the flow scales to $\langle{u}_z(\bm{x}^p(t),t)\rangle$. Second, we will study the effect of 2WC at different scales and also look at the relative contribution of different scales to the overall settling velocity enhancement. Third, we will analyze the effect of the two contributions embedded in \eqref{Alt} that affects settling velocity at different scales and investigate how these are modified by 2WC at different scales of the flow. Thus, we aim to elucidate how 2WC modifies the multiscale preferential sweeping mechanism.

\section{Direct Numerical Simulations}\label{sec:DNS_Details}
\begin{table}
\begin{center}
\begin{tabular}{c c c c c c c}
Simulation                  & \qquad I             & \qquad II            & \qquad III           & \qquad IV           & \qquad V           \\
$Re_{\lambda}$               & \qquad 31            & \qquad 54            & \qquad 75            & \qquad 87           & \qquad 142         \\
$\mathcal{L}$               & \qquad $2\pi$        & \qquad $2\pi$        & \qquad $2\pi$        & \qquad $2\pi$       & \qquad $2\pi$      \\
$N$                         & \qquad 32            & \qquad 64            & \qquad 96            & \qquad 128          & \qquad 256         \\
$\nu$                       & \qquad 0.032         & \qquad 0.013         & \qquad 0.007         & \qquad 0.005        & \qquad 0.002       \\
$\langle\epsilon\rangle$    & \qquad 0.336         & \qquad 0.279         & \qquad 0.271         & \qquad 0.279        & \qquad 0.263       \\
$dt$                        & \qquad 0.005         & \qquad 0.005         & \qquad 0.003         & \qquad 0.002        & \qquad 0.001       \\
$L$                         & \qquad 1.870         & \qquad 1.652         & \qquad 1.511         & \qquad 1.431        & \qquad 1.417       \\
$u'$                        & \qquad 0.916         & \qquad 0.916         & \qquad 0.916         & \qquad 0.916        & \qquad 0.916       \\
$\tau_L$                    & \qquad 2.041         & \qquad 1.803         & \qquad 1.649         & \qquad 1.562        & \qquad 1.547       \\
$\eta$                      & \qquad 0.100         & \qquad 0.053         & \qquad 0.034         & \qquad 0.026        & \qquad 0.013       \\
$u_{\eta}$                  & \qquad 0.322         & \qquad 0.245         & \qquad 0.209         & \qquad 0.193        & \qquad 0.151       \\
$\tau_{\eta}$               & \qquad 0.309         & \qquad 0.216         & \qquad 0.161         & \qquad 0.134        & \qquad 0.087       \\
$\mathcal{L}/L$             & \qquad 3.36          & \qquad 3.80          & \qquad 4.16          & \qquad 4.39         & \qquad 4.43        \\
$L/\eta$                    & \qquad 18.79         & \qquad 31.16         & \qquad 45.10         & \qquad 55.24        & \qquad 107.33      \\
$u'/u_{\eta}$               & \qquad 2.85          & \qquad 3.74          & \qquad 4.39          & \qquad 4.75         & \qquad 6.05        \\
$\tau_L/\tau_{\eta}$        & \qquad 6.60          & \qquad 8.33          & \qquad 10.26         & \qquad 11.63        & \qquad 17.74       \\
$k_{max}\eta$               & \qquad 1.50          & \qquad 1.60          & \qquad 1.52          & \qquad 1.56         & \qquad 1.59        \\
$N_{proc}$              & \qquad 16                & \qquad 16            & \qquad 16            & \qquad 64           & \qquad 256         \\
\end{tabular}

\caption{Flow parameters in DNS of unladen isotropic turbulence where all dimensional parameters are in arbitrary units. $Re_{\lambda} \equiv u'\lambda/\nu = 2k/\sqrt{5/3\nu \langle\epsilon\rangle} $ is the Taylor microscale Reynolds Number,  $\lambda$ is the Taylor microscale, $\mathcal{L}$ is the domain length, $N$ is the number of grid points in each direction, $\nu$ is the fluid kinematic viscosity, $\langle\epsilon\rangle$ is the mean turbulent kinetic energy dissipation rate, $L$ is the integral length scale, $\eta \equiv (\nu^{3}/\langle\epsilon\rangle)^{1/4} $ is the Kolmogorov length scale, $u' \equiv \sqrt{2k/3}$ is the r.m.s of fluctuating fluid velocity, $k$ is the turbulent kinetic energy, $u_{\eta}$ is the Kolmogorov velocity scale, $\tau_L \equiv L/u'$ is the large-eddy turnover time, $\tau_{\eta}$ is the Kolmogorov time scale, $k_{max} = \sqrt{2N/3}$ is the maximum resolved wavenumber, $dt$ is the time step  and $N_{proc}$ is the number of processors used for the simulations. The small-scale resolution, $k_{max}\eta$ and the total flow kinetic energy measured by $u'$ are approximately constant between the different simulations. All statistics are averaged over the last 10 $\tau_L$ of the total run time of 20 $\tau_L$ for the unladen simulations.}
\label{Table_Unladen}

\end{center}
\end{table}

We simulate numerically a homogeneous turbulent flow laden with settling inertial particles. The fluid flow is discretized on a Cartesian grid through an Eulerian approach, while a Lagrangian particle tracking approach is used for the particles. For consistency, the same point-particle equation of motion \eqref{eqn_eom} is used as in the theory, and a major numerical challenge is to couple the fluid and particulate phases through their momentum exchange.

\begin{table}
\begin{center}
\setlength{\tabcolsep}{6pt}
\begin{tabular}{lcccccccccc}
$St$                       & \multicolumn{2}{c}{0.3}                           &  \multicolumn{2}{c}{0.5}                           &  \multicolumn{2}{c}{0.7}                           & \multicolumn{2}{c}{1.0}                           &  \multicolumn{2}{c}{2.0}                           \\
$N_p$                      & \multicolumn{2}{c}{11,506,176}                    & \multicolumn{2}{c}{5,347,840}                     & \multicolumn{2}{c}{3,228,672}                     & \multicolumn{2}{c}{1,890,816}                     & \multicolumn{2}{c}{668,672}                       \\
$d_p (\times 10^{-3})$     & \multicolumn{2}{c}{0.85}                          & \multicolumn{2}{c}{1.10}                          & \multicolumn{2}{c}{1.30}                          & \multicolumn{2}{c}{1.55}                          & \multicolumn{2}{c}{2.20}                          \\
$d_p / \Delta x$           & \multicolumn{2}{c}{0.017}                         & \multicolumn{2}{c}{0.022}                         & \multicolumn{2}{c}{0.027}                         & \multicolumn{2}{c}{0.032}                         & \multicolumn{2}{c}{0.045}                         \\
Coupling                   & \multicolumn{1}{l}{1WC} & \multicolumn{1}{l}{2WC} & \multicolumn{1}{l}{1WC} & \multicolumn{1}{l}{2WC} & \multicolumn{1}{l}{1WC} & \multicolumn{1}{l}{2WC} & \multicolumn{1}{l}{1WC} & \multicolumn{1}{l}{2WC} & \multicolumn{1}{l}{1WC} & \multicolumn{1}{l}{2WC} \\
$Re_{\lambda}$             & 89.2                    & 88.3                    & 89.4                    & 88.8                    & 89.3                    & 89.3                    & 89.4                    & 92.1                    & 91.1                    & 95.1                    \\
$\langle \epsilon \rangle$ & 0.266                   & 0.272                   & 0.266                   & 0.268                   & 0.266                   & 0.266                   & 0.265                   & 0.250                   & 0.256                   & 0.235                   \\
$L$                        & 1.478                   & 1.484                   & 1.479                   & 1.481                   & 1.472                   & 1.487                   & 1.475                   & 1.518                   & 1.506                   & 1.505                   \\
$\eta$                     & 0.026                   & 0.026                   & 0.026                   & 0.026                   & 0.026                   & 0.026                   & 0.026                   & 0.027                   & 0.027                   & 0.027                   \\
$\tau_{\eta}$              & 0.137                   & 0.136                   & 0.138                   & 0.137                   & 0.137                   & 0.137                   & 0.138                   & 0.142                   & 0.140                   & 0.146                   \\
$u_{\eta}$                 & 0.191                   & 0.192                   & 0.191                   & 0.191                   & 0.191                   & 0.191                   & 0.191                   & 0.188                   & 0.189                   & 0.185                   \\
$\tau_L$                   & 1.613                   & 1.620                   & 1.614                   & 1.617                   & 1.607                   & 1.623                   & 1.610                   & 1.657                   & 1.644                   & 1.643                  
\end{tabular}

\caption{Flow parameters in the DNS of particle-laden homogeneous turbulence for the one-way coupled (1WC) and two-way coupled (2WC) regimes in arbitrary units, and for varying Stokes number, $St$. $d_p$ is the particle diameter and $\Delta x$ is the grid spacing. See table \ref{Table_Unladen} for the definition of other parameters. All the simulations in this table correspond to unladen simulation IV ($N =128$, $Re_{\lambda} = 87$) in table \ref{Table_Unladen} prior to the introduction of the particles. Note that $St$ is defined based on Kolmogorov parameters in the unladen flow. These simulations used density ratio $\rho_p/\rho_f = 5000$,  Froude number $\Fr = 1$, and volume fraction $\Phi_v = 1.5 \times 10^{-5}$. All statistics are averaged over the last 100 $\tau_L$ of the total run time of 120 $\tau_L$ of both the 1WC and 2WC simulations.}
\label{Table_StDep}

\end{center}
\end{table}

\subsection{Fluid solver}
We solve the three-dimensional, incompressible Navier-Stokes equation in rotational form
\begin{align}
\partial_t \vect{u}  =
\vect{u}\times\vect{\omega}
-\bnabla\left(\frac{P}{\rho_f} + \frac{u^2}{2}\right) + \nu\nabla^2\vect{u} + \vect{f} + \vect{C},\quad \bnabla\bcdot\vect{u} = 0
\label{eq_NS}
\end{align}
using the HiPPSTR code \citep{ireland13}, modified to account for the particle feedback on the flow $\vect{C}$. Here $\vect{\omega}\equiv\bnabla\times\vect{u}$ is the vorticity, $P$ is the pressure, $\rho_f$ is the constant density of the fluid, $\nu$ is the kinematic viscosity, $\vect{f}$ is a large-scale forcing and $\vect{C}$ is the force exerted by the particles on the fluid. 
The spatial derivatives in equations \eqref{eq_NS} are discretized through a Fourier pseudo-spectral method. The non linearity introduces an aliasing error that is removed by using a combination of spherical truncation and phase shifting.
The pressure gradient compensates the divergence of both the non-linear convective term and the divergence of $\vect{C}$, thus yielding a solenoidal fluid velocity field.
The Fourier modes are evolved in time by means of a second-order Runge-Kutta scheme with exponential integration for the viscous stress. A large-scale deterministic forcing scheme is applied at wavenumbers with magnitude $\kappa \leq \sqrt{2}$ that maintains a constant kinetic energy of the flow.

\begin{table}
\begin{center}
\setlength{\tabcolsep}{6pt}
\begin{tabular}{lllllllllll}
$N$                        & \multicolumn{2}{c}{32}     & \multicolumn{2}{c}{64}      & \multicolumn{2}{c}{96}      & \multicolumn{2}{c}{128}       & \multicolumn{2}{c}{256}        \\
$N_p$                      & \multicolumn{2}{c}{33,408} & \multicolumn{2}{c}{220,672} & \multicolumn{2}{c}{871,296} & \multicolumn{2}{c}{1,890,816} & \multicolumn{2}{c}{14,284,800} \\
$d_p (\times 10^{-3})$     & \multicolumn{2}{c}{5.97}   & \multicolumn{2}{c}{3.18}    & \multicolumn{2}{c}{2.01}    & \multicolumn{2}{c}{1.55}      & \multicolumn{2}{c}{0.79}       \\
$d_p / \Delta x$           & \multicolumn{2}{c}{0.030}  & \multicolumn{2}{c}{0.032}   & \multicolumn{2}{c}{0.031}   & \multicolumn{2}{c}{0.032}     & \multicolumn{2}{c}{0.032}      \\
Coupling                   & 1WC          & 2WC         & 1WC           & 2WC         & 1WC           & 2WC         & 1WC           & 2WC           & 1WC            & 2WC           \\
$Re_{\lambda}$             & 31.9         & 31.3        & 53.9          & 53.9        & 74.8          & 77.0        & 89.4          & 92.1          & 144.9          & 149.9         \\
$\langle \epsilon \rangle$ & 0.330        & 0.340       & 0.281         & 0.281       & 0.271         & 0.255       & 0.265         & 0.250         & 0.253          & 0.237         \\
$L$                        & 1.874        & 1.866       & 1.646         & 1.645       & 1.518         & 1.553       & 1.475         & 1.518         & 1.441          & 1.463         \\
$\eta$                     & 0.100        & 0.099       & 0.053         & 0.053       & 0.034         & 0.034       & 0.026         & 0.027         & 0.013          & 0.014         \\
$\tau_{\eta}$              & 0.314        & 0.308       & 0.216         & 0.216       & 0.161         & 0.166       & 0.138         & 0.142         & 0.089          & 0.092         \\
$u_{\eta}$                 & 0.319        & 0.322       & 0.246         & 0.246       & 0.208         & 0.205       & 0.191         & 0.188         & 0.150          & 0.147         \\
$\tau_L$                   & 2.045        & 2.037       & 1.796         & 1.796       & 1.657         & 1.696       & 1.610         & 1.657         & 1.573          & 1.597        
\end{tabular}

\caption{Flow parameters in the DNS of particle-laden homogeneous turbulence for the one-way coupled (1WC) and two-way coupled (2WC) regimes in arbitrary units, and for varying Taylor Reynolds number, $Re_{\lambda}$. $d_p$ is the particle diameter and $\Delta x$ is the grid spacing. See table \ref{Table_Unladen} for the definition of other parameters. All the simulations in this table correspond to unladen simulation with corresponding $N$ in table \ref{Table_Unladen} prior to the introduction of the particles.  These simulations used density ratio $\rho_p/\rho_f = 5000$,  Froude number $\Fr = 1$, Stokes number $\St = 1$, and volume fraction $\Phi_v = 1.5 \times 10^{-5}$. Note that $St$ is defined based on Kolmogorov parameters in the unladen flow. All statistics are averaged over the last 100 $\tau_L$ of the total run time of 120 $\tau_L$ of both the 1WC and 2WC simulations.}
\label{Table_ReDep}

\end{center}
\end{table}

\subsection{Particle solver}

The particle position and velocity are evolved using \eqref{eqn_eom}, with the fluid velocity at the particle position computed through B-spline interpolation \citep{vanhinsberg12,carbone2018}. This operation is understood as a backward Non-Uniform Fast Fourier Transform (NUFFT) with B-spline basis \citep{beylkin95}. The fluid velocity field is projected onto the B-spline basis in Fourier space, then brought back to physical space by means of an inverse Fast Fourier Transform (FFT). Finally, this projected velocity field is interpolated at the particle position. We employed B-spline polynomials of order seven, which uses 8 points for interpolation along each direction and has $C^6$ continuity \citep{ireland16a}.
The particles position and velocity are then evolved in time by means of an exponential integrator \citep{hochbruck2010}, that guarantees stability and accuracy even for very low Stokes numbers. For a more complete description of the DNS solver see \cite{ireland13}. Note, however, that in the current version of the HiPPSTR code used to simulate the 2WC system, the same exponential integration scheme is used for time advancement of both the fluid and particle solvers (unlike the original HiPPSTR code which used exponential integrators only for the particles), to ensure consistency between the handling of the two phases. 

\subsection{Flow Parameters}

The particles simulated are monodisperse, having the same radius and density. In order to be able to compare our results with those of \cite{monchaux17} we use the same density ratio that they did, namely $\rho_p/\rho_f=5000$, the same Froude number, $\Fr=1$, and the same volume fraction, $\Phi_v=1.5\times 10^{-5}$, corresponding to a mass fraction
$\Phi_m= 0.075$.
We consider Stokes numbers $\St=0.3, 0.5, 0.7, 1, 2 $ in order to explore the behavior in the regimes of weak to moderate inertia. With these values of $\Fr$ and $\St$, the artificial periodicity effects discussed in \cite{ireland16b} do not arise, thus avoiding the need to use very large flow domains in the direction of the gravity. 

It is also desirable to consider the effects of the Taylor Reynolds number $Re_\lambda$ on the flow, since as shown in \cite{tom19}, the range of scales in the flow is of crucial importance in determining the particle settling behavior via the multiscale preferential sweeping mechanism. We found that it was necessary to run the DNS for very long times, on the order of 100 eddy turnover times, in order for the average vertical velocities of the particles to statistically converge. This then places severe limitations on the $Re_\lambda$ that can be simulated (due to limited computational resources), and as a result, we considered the range $Re_\lambda\in [31,142]$. Although this range is too small to observe how the settling process might approach an asymptotic state with increasing $Re_\lambda$ for sufficiently large $Re_\lambda$ (as suggested in \cite{tom19}), it will nevertheless be shown to be sufficient to reveal significant effects of $Re_\lambda$ on the settling process.

\subsection{Two-way coupling}

In the framework of the point-particle model, the force from the particles on the fluid is a superposition of Dirac delta functions centered at the particle position
\begin{equation}
\vect{C}\of{\vect{x},t} = \frac{L^3}{N_p}\Phi_m \sum_p
\frac{\vect{v}^p(t) - \vect{u}\of{\vect{x}^p,t}}{\tau_p}
 \delta(\vect{x} - \vect{x}^p(t)).
\label{eq_coupling_C}
\end{equation}
Note that in this term, only the drag force on the particle is accounted for and not the gravitational body force experienced by the particle, as was also done in previous works (e.g. \cite{bosse06,monchaux17}). The reason for this is that this term is supposed to be equal and opposite to the force that the fluid applies to the particles located at $\vect{x}$, and the latter is associated with the hydrodynamic stresses on the particle surfaces (represented by the drag force in the point-particle model being used) and not with the gravitational body force.

The Fourier transform of \eqref{eq_coupling_C} needed by the pseudo-spectral solver for the fluid phase is computed by means of the forward Non-Uniform Fast Fourier Transform (NUFFT) with B-spline basis \citep{beylkin95,carbone2018,carbone2019}. 
This is motivated by the fact that a Fast Fourier Transform (FFT) is not directly applicable to \eqref{eq_coupling_C}, while a direct Discrete Fourier Transform (DFT) would be computationally too expensive.
A detailed description of the NUFFT algorithm for particles in turbulence can be found in \cite{carbone2018}.
Briefly, the algorithm can be summarized through the equation
\begin{equation}
\mathcal{F}[\vect{C}] = \frac{\mathcal{F}[\vect{C}*B]}{\mathcal{F}[B]},
\end{equation}
where $\mathcal{F}$ indicates the Fourier transform, $*$ denotes a convolution and $B$ is the B-spline polynomial.
First, we compute the convolution of the coupling term \eqref{eq_coupling_C} with the B-spline polynomial basis $B\of{\vect{x}}$ in physical space. This smoothing allows representing the coupling term on the structured Cartesian grid. The smoothed field is then transformed to Fourier space by means of a standard FFT. As a last step, the initial convolution with the B-spline is removed in Fourier space by dividing the Fourier transform of the coupling by the Fourier transform of the B-spline polynomial (i.e.~a power of the sinc function).
After these three steps we obtain the contribution from the particles to the time evolution of the fluid velocity Fourier modes.
The same method has been recently used to investigate the thermal coupling between a passive temperature field and inertial particles \citep{carbone2019}, and the results presented in this work have been crosschecked with the results from the code used in \cite{carbone2019}.

Most previous works used tri-linear extrapolation to distribute the contribution of the particle on the nearest grid points \citep{bosse06}. One of the main reasons for this choice is that a low order polynomial reduces the artificial non localities that arise due to spreading the effect of the particle momentum onto the surrounding grid points. In the NUFFT, the initial convolution in physical space is subsequently removed in Fourier space, so that the method does not suffer from the artificial non-locality issues, which is a major advantage of the method. As a consequence, this approach allows to use higher-order B-spline polynomials to obtain a more accurate representation of the Dirac delta functions in \eqref{eq_coupling_C}. Moreover, many recent works \citep[e.g.][]{rosa21} used high-order schemes for the interpolation of the fluid velocity at the particle position combined with low-order schemes for the computation of the coupling term. Since this introduces errors in the energy balance of the system \citep{sundaram96}, we employ the same polynomial basis (of degree seven) for interpolation of the velocity field and computation of the coupling term $\vect{C}$.

\subsection{Numerical considerations}

In the particle equation of motion \eqref{eqn_eom}, $\bm{u}(\bm{x}^p(t),t)$ is the undisturbed fluid velocity at the particle position, meaning the fluid velocity that does not include the disturbance of the particle under consideration (but can include the disturbance from all other particles in the flow). As a result, simply interpolating the fluid velocities from the surrounding grid points to evaluate $\bm{u}(\bm{x}^p(t),t)$ introduces an error because the interpolated velocities include the disturbance effect of the particle under consideration. This error can be significant for particles of size comparable to the grid spacing employed in the numerical simulation, i.e. when $d_p=O(\Delta x)$ \citep{boivin98}. Moreover, this issue is particularly evident for particles settling in turbulence, and even in a still fluid \citep{horwitz2020}. Correction schemes have been recently developed to approximately remove this error by providing a way to retrieve the undisturbed fluid velocity \citep{horwitz2018}. However, in the regime $d_p\ll\Delta x$ the error will be small, and in our simulations $d_p/\Delta x \leq 0.05$. As a result, a correction scheme is not required, and we simply evaluate $\bm{u}(\bm{x}^p(t),t)$ in \eqref{eqn_eom} by interpolating the fluid velocities at the surrounding grid point to the particle position, as described earlier.

Finally, following previous works \citep{bosse06,monchaux17,rosa21}, at each time-step we set to zero the average fluid velocity in the direction of gravity. This is equivalent to applying a mean vertical pressure gradient \citep{bosse06,monchaux17,rosa21} without which the total kinetic energy in the flow would grow without bound due to the finite settling velocity of the particles continuously injecting kinetic energy into the flow. While introduced in \citet{maxey01} and used for 2WC simulations by \citet{bosse06} to remedy the numerical issue that arises for an unbounded flow, its physical validity could be called into question. However, a possible justification can be given as follows. Suppose one is considering the settling of heavy particles through a fully developed wall-bounded turbulent flow with 2WC at distances from the wall where the flow is quasi-homogeneous. In this case, although the settling particles will continuously force the flow in the vertical direction as they fall, the impermeability of the wall prevents a mean flow from developing in this direction, and this effect is mediated to the upper, quasi-homogeneous region of the flow through the mean forces in the fluid. Hence in this situation, the settling particles would not generate a mean flow as they settle (although the mean fluid velocity evaluated at the particle positions can still be finite, and indeed will be if the flow modifies the particle settling velocity compared to the Stokes settling velocity). This potential justification is also hinted at in \citet{maxey01}. In addition to this potential justification, we also use this aforementioned method of enforcing a zero mean fluid velocity at each time step in order to be able to meaningfully compare our results with those of the previous studies by \cite{bosse06,monchaux17,rosa21}.

\subsection{Simulation approach and post-processing methodology}
We found that very long simulation times are necessary in order to obtain converged statistics of the average vertical fluid velocity at the inertial particle position and therefore the particle settling velocity, a requirement that was also mentioned in \citet{bosse06} and \citet{rosa16}. As such, our statistics are computed over the very long time window of 100 large-scale eddy turnover times. We ran the fluid only simulation for 20 $\tau_L$ and obtained the flow statistics by averaging over the last 10 $\tau_L$ (see table \ref{Table_Unladen}). Next, we introduced particles into the flow and ran the 1WC case for 120 $\tau_L$ and computed the 1WC statistics over the the last 100 $\tau_L$. Finally, we started the 2WC simulation from the instantaneous velocity and particle field of
the completed 1WC run to obtain faster convergence of particle and fluid statistics and also because using the 1WC particle
initial conditions in the 2WC regime do not affect
the long-term statistics \citep{bosse06}.  Similar to the 1WC case, the 2WC case was also ran for 120 $\tau_L$ and the 2WC statistics were computed over the the last 100 $\tau_L$. Some of the previous works have used small averaging windows, (for e.g. \citet{good14} uses 10 $\tau_L$) and as pointed out by \citet{rosa16}, this could be potential reason for different quantitative results between various works. The fluid statistics can change once the two-way coupling is activated and thus the 2WC fluid statistics are only known \textit{a posteriori}. As is common in 2WC studies, the particle Stokes number $St$ in the 1WC and 2WC simulations are computed based on fluid statistics of the unladen flow presented in table \ref{Table_Unladen}.

We investigate the contribution to the particle settling from different scales by computing relevant quantities filtered at various scales. We employ a sharp filter in Fourier space that filters out all the modes with wavenumber magnitude larger than the characteristic wavenumber $k_F$, which is associated to a filtering length $\ell_F=2\pi/k_F$ \citep{eyink09} and obtain the sub-grid field through
\begin{align}
{u}'_z(\bm{x},t)\equiv u_z(\bm{x},t)-\widetilde{u}_z(\bm{x},t)=\Large\sum_{\|\bm{k}\|> k_F}\hat{u}_z(\bm{k},t)e^{i\bm{k\cdot x}},
\end{align}
where $\widetilde{(\cdot)}$ denotes the coarse-grained field, while $(\cdot)'$ denotes the sub-grid field. We then interpolate $u'_z(\bm{x},t)$ to the positions of the inertial particles $\bm{x}^p(t)$ using an eight-point B-spline interpolation scheme to obtain ${u}'_z(\bm{x}^p(t),t)$. The values of ${u}'_z(\bm{x}^p(t),t)$ are then averaged over all the particles and over multiple times to obtain $\langle{{u}'_z}(\bm{x}^p(t),t)\rangle$. This process is then repeated for multiple $k_F$ in order to examine the contribution of the different flow scales. This is the same approach that was adopted in our previous paper \citep{tom19} for analyzing the multiscale preferential sweeping mechanism.

\section{Results and discussion}\label{sec:DNS_Results}

\FloatBarrier

\subsection{Unfiltered results}

\begin{figure}
\centering
\vspace{0mm}			
    \subfloat[]	
	{\begin{overpic}
	[trim = 0mm -25mm -40mm -1mm,
	scale=0.125,clip,tics=20]{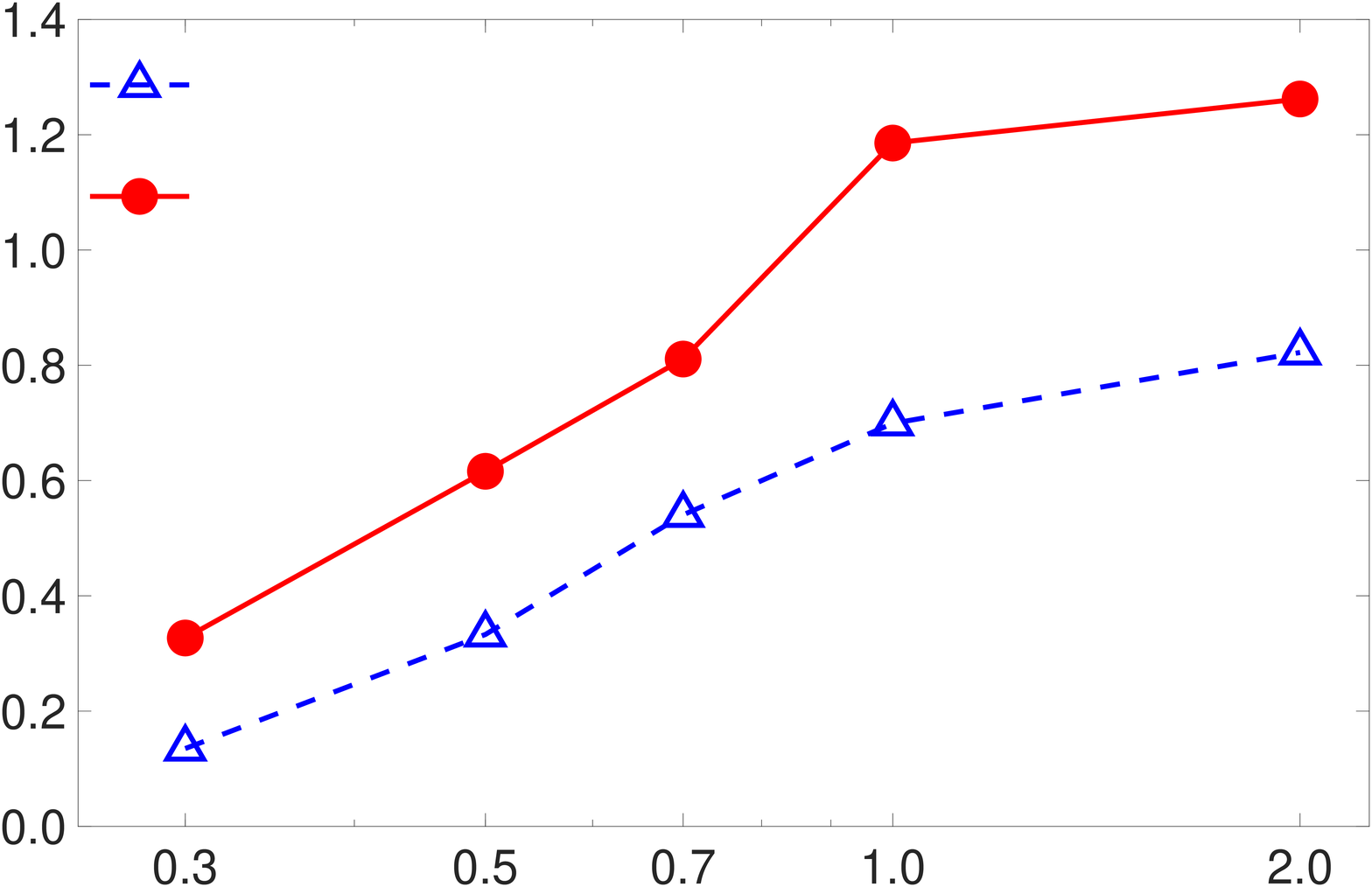}
	    \put(54,0){$St$}
        \put(-8,16){\rotatebox{90}{$-\langle u_z(\bm{x}^p(t),t)\rangle / u_{\eta}$}}
        \put(14.5,57.75){1WC}
        \put(14.5,49.75){2WC}
	\end{overpic}}
	\subfloat[]
	{\begin{overpic}
	[trim = -10mm -25mm 0mm -1mm,
	scale=0.125,clip,tics=20]{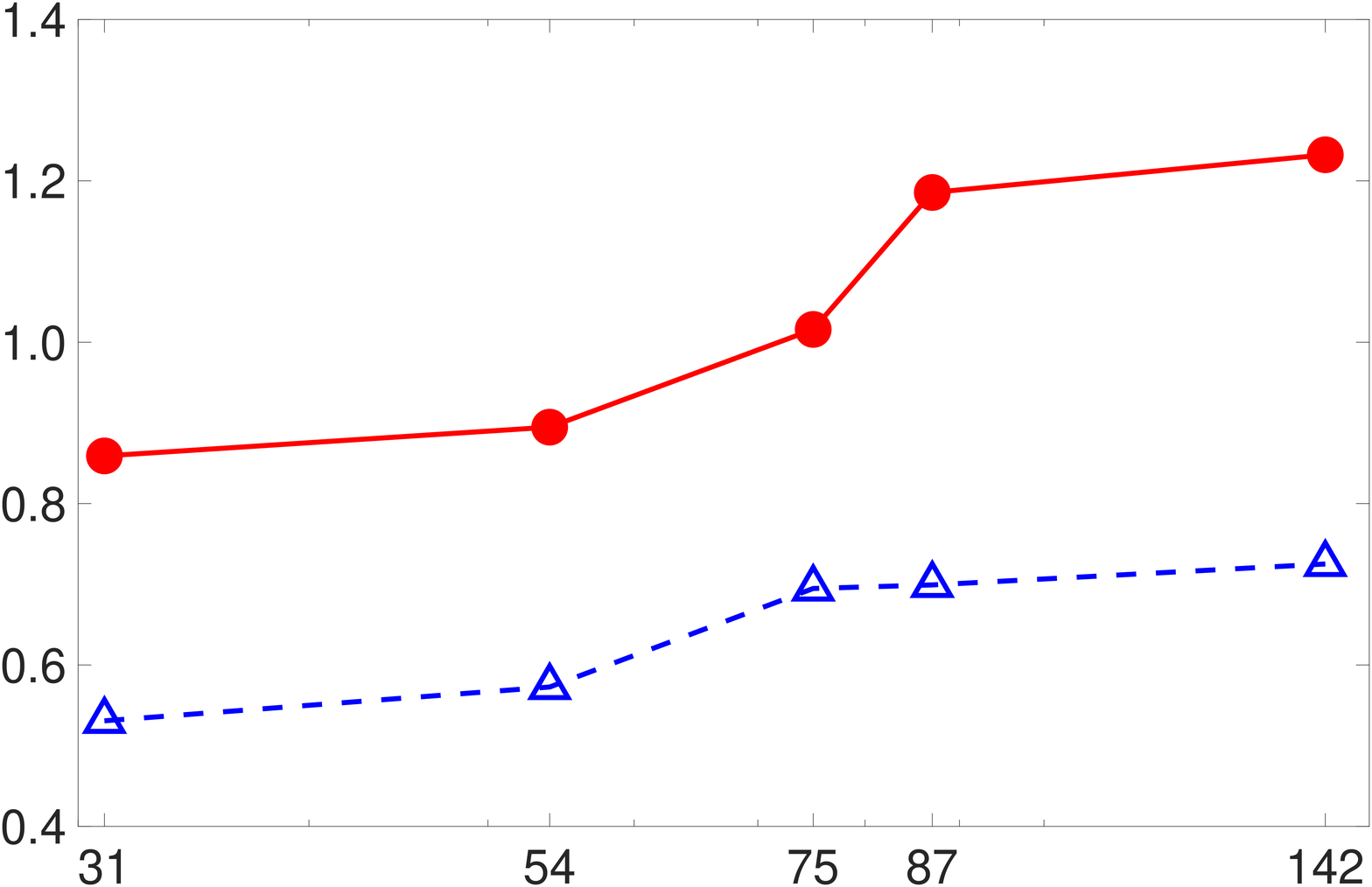}
	    \put(54,0){$Re_{\lambda}$}
	\end{overpic}} 
	\caption{Normalized settling velocity enhancement for the 1WC and 2WC cases shown as a function of (a) $St$ and (b) $Re_{\lambda}$.
	}
    \label{Unfilt_SGSVel}
\end{figure}

We begin by considering the unfiltered results, which includes the effects of all scales in the flow on the particle settling. As discussed in \S\ref{Background}, the average particle velocity may differ from the Stokes settling velocity only if $\langle{u}_z(\bm{x}^p(t),t)\rangle\neq{0}$ and it is therefore this quantity that we focus on in order to understand how the turbulence modifies the particle settling, and how this is influenced by 2WC. We will refer to $\langle{u}_z(\bm{x}^p(t),t)\rangle$ as the `full' settling velocity enhancement in subsequent discussions as it includes the effect of all the flow scales. 

In figures \ref{Unfilt_SGSVel} (a) and (b), we show $-\langle{u}_z(\bm{x}^p(t),t)\rangle/u_\eta$ for both the one-way and 2WC regimes for a range of $St$ and $Re_{\lambda}$, respectively, and for the fixed values $Fr = 1$ and $\Phi_v = 1.5 \times 10^{-5}$. The results show that $-\langle{u}_z(\bm{x}^p(t),t)\rangle/u_\eta$ is significantly larger for the 2WC case compared to the 1WC case, as reported in previous works \citep{bosse06,monchaux17,rosa21}. It is worth emphasizing that we see this significant enhancement even for the low volume fraction considered in this study $\Phi_v=1.5\times 10^{-5}$, where the effect of 2WC on the global flow statistics is negligible (see table \ref{Table_StDep} and \ref{Table_ReDep}). This suggests, in agreement with the previous studies of \cite{bosse06,monchaux17,rosa21}, that in modeling atmospheric particle transport where $\Phi_v$ is often small, the effects of 2WC on particle settling may not be negligible.

Next, we compare the quantitative values of the settling enhancement to other studies in table \ref{Table_Comapre}. The study by \cite{rosa21} focuses on the microphysical processes relevant for cloud droplets in typical atmospheric flows and hence uses a density ratio $\rho_p / \rho_f = 1000$ and $Fr$ values relevant to those particular problems. Therefore, we only compare our results with the works of \citet{bosse06} and \citet{monchaux17}, who use the same density ratio $\rho_p / \rho_f = 5000$  and volume fraction $\Phi_v=1.5\times 10^{-5}$ considered in this work. While all the works referred to in table \ref{Table_Comapre} show that 2WC strongly enhances the settling compared with the 1WC case, there are some differences in the values of the enhancement reported in these studies. We discuss four potential reasons for this discrepancy. First, the work by \citet{monchaux17} used a particle-in-cell (PIC) method for distributing the feedback force from the particle locations to the surrounding grid points (`feedback spreading'). While PIC is a popular method, its accuracy has been called into question because the coupling term is strongly grid dependent unless the number of particles per cell exceeds a certain threshold. \citep{balachandar09,garg09,gualtieri13,gualtieri15}  Second, \citet{monchaux17} used different methods for interpolation and feedback spreading and this can introduce errors into the energy balance of the system \citep{sundaram96}. Third, \citet{bosse06} made use of the `computational particles' approach \citep{elghobashi94} (also referred to as the `super-droplet approach'), which on the one hand reduces the computational costs at larger mass loading, but on the other hand, can affect the accuracy of the results. Fourth, as was demonstrated in \cite{tom19}, although the settling velocity enhancement depends on a restricted range (determined by $St$) of scales of the flow, this range could span the entire range of scales in the flows simulated in these DNS studies due to the low values of $Re_{\lambda}$. The range of velocity scales available in the flow as characterized by $u'/u_{\eta}$ is different for all the three studies in table \ref{Table_Comapre} and that could be another cause of the quantitative differences. However, although there are quantitative differences between our study and those of \citet{bosse06} and \cite{monchaux17}, we emphasize that our results are comparable to theirs, with each study demonstrating that 2WC can strongly enhance particle settling speeds in homogeneous turbulent flows.

\begin{table}
\begin{center}
\setlength{\tabcolsep}{4pt}
\begin{tabular}{cccc}
\textbf{Study}                   & \textbf{Present}   & \textbf{\cite{bosse06}}     & \textbf{\cite{monchaux17}}          \\
\hline
\textit{\textbf{Interpolation scheme}}    & $7^{th}$ order B-Spline & Tri-linear interpolation & $4^{th}$ order Lagrangian Polynomial \\
\textit{\textbf{Feedback spreading}}      & $7^{th}$ order B-Spline & Tri-linear interpolation & Particle-in-Cell (PIC) method   \\
\textit{\textbf{Computational particles}} & No                 & Yes                     & No                              \\
\textit{\textbf{$\rho_p / \rho_f$}}       & 5000               & 5000                    & 5000                            \\
\textit{\textbf{$N$}}                     & 64                 & 64                      & 64                              \\
\textit{\textbf{$Re_{\lambda}$}}          & 54                 & 43                      & 40                              \\
\textit{\textbf{$u' / u_{\eta}$}}         & 3.74               & 3.30                    & 1.99                            \\
\textit{\textbf{$1WC \quad - \Delta v / u'  $ }}   & 0.15              & 0.11                   & 0.08                           \\
\textit{\textbf{$2WC \quad - \Delta v / u'  $ }}   & 0.24              & 0.18                   & 0.22                          
\end{tabular}

\caption{Comparison of settling velocity enhancement measured by $\Delta v \equiv \langle{v}_z^p(t)\rangle + \tau_p{g}$ for one-way (1WC) and two-way (2WC) coupled flows in various studies for $\rho_p / \rho_f = 5000$ and and volume fraction, $\Phi_v=1.5\times 10^{-5}$. The results presented are for $St = 1$ and $Sv = 1$ (and hence a $Fr \equiv St / Sv = 1$). The results from \cite{bosse06} are based on table II (corresponding to Runs 0 and 2) and results from \cite{monchaux17} are based on figure 2(a) (corresponding to Rouse number, $R \equiv St u_{\eta} / Fr u' \approx 0.5$) in the corresponding manuscripts.}
\label{Table_Comapre}

\end{center}
\end{table}

\begin{figure}
\centering
\vspace{0mm}			
	{\begin{overpic}
	[trim = 0mm -25mm -55mm -3mm,
	scale=0.2,clip,tics=20]{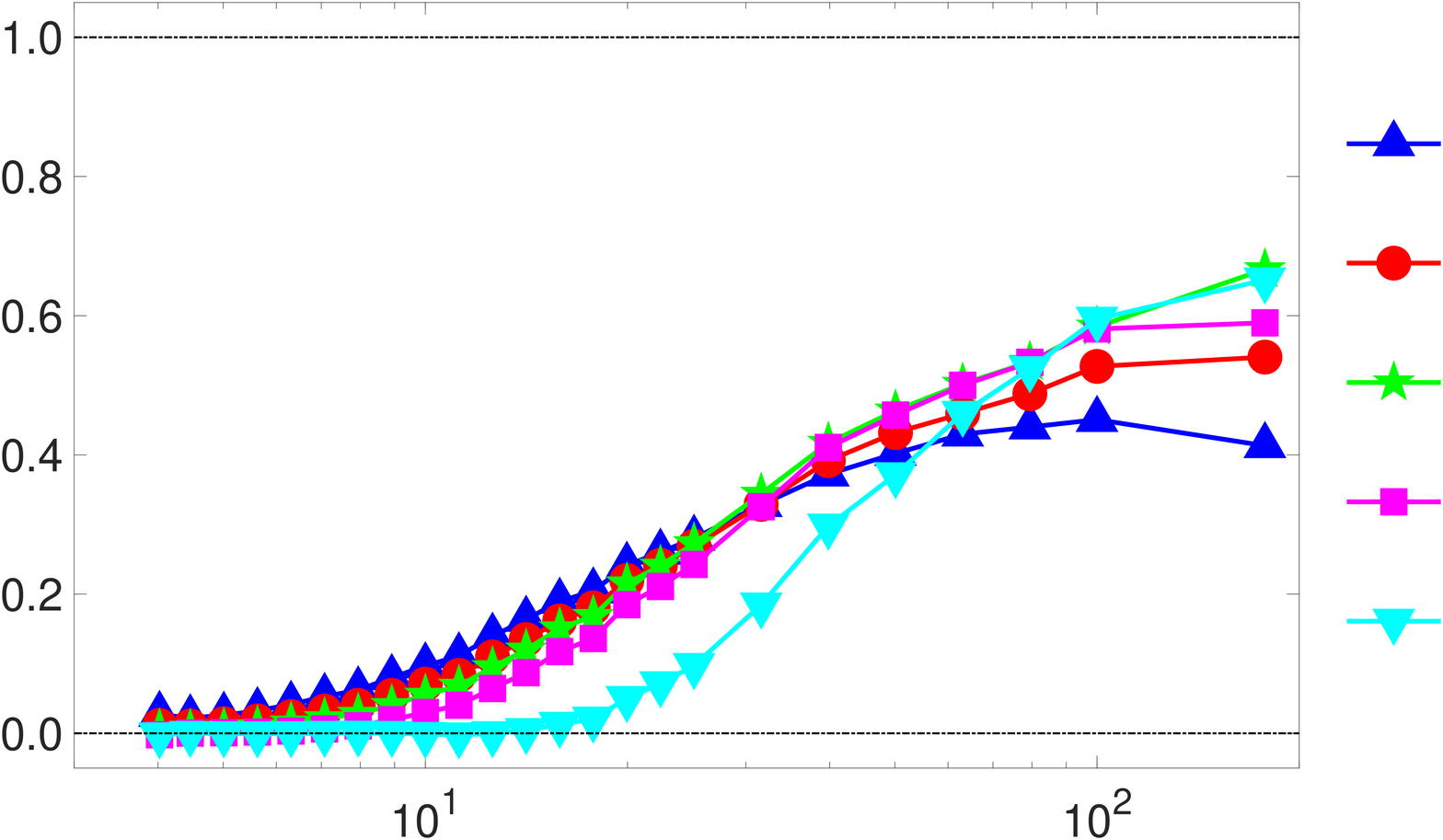}
\put(45,1){\large{$\ell_F / \eta$}}
\put(-10,22){\rotatebox{90}{\Large{$ \frac{ \langle u'_{z,1WC}(\bm{x}^p(t),t)\rangle } {\langle u'_{z,2WC}(\bm{x}^p(t),t)\rangle} $}}}
\put(92,47.0){\large{$St = 0.3$}}
\put(92,39.5){\large{$St = 0.5$}}
\put(92,32.0){\large{$St = 0.7$}}
\put(92,24.5){\large{$St = 1.0$}}
\put(92,17.0){\large{$St = 2.0$}}
\put(5,26){
\begin{overpic}[trim = 0mm 0mm 0mm 0mm, scale=0.10,clip,tics=20]
{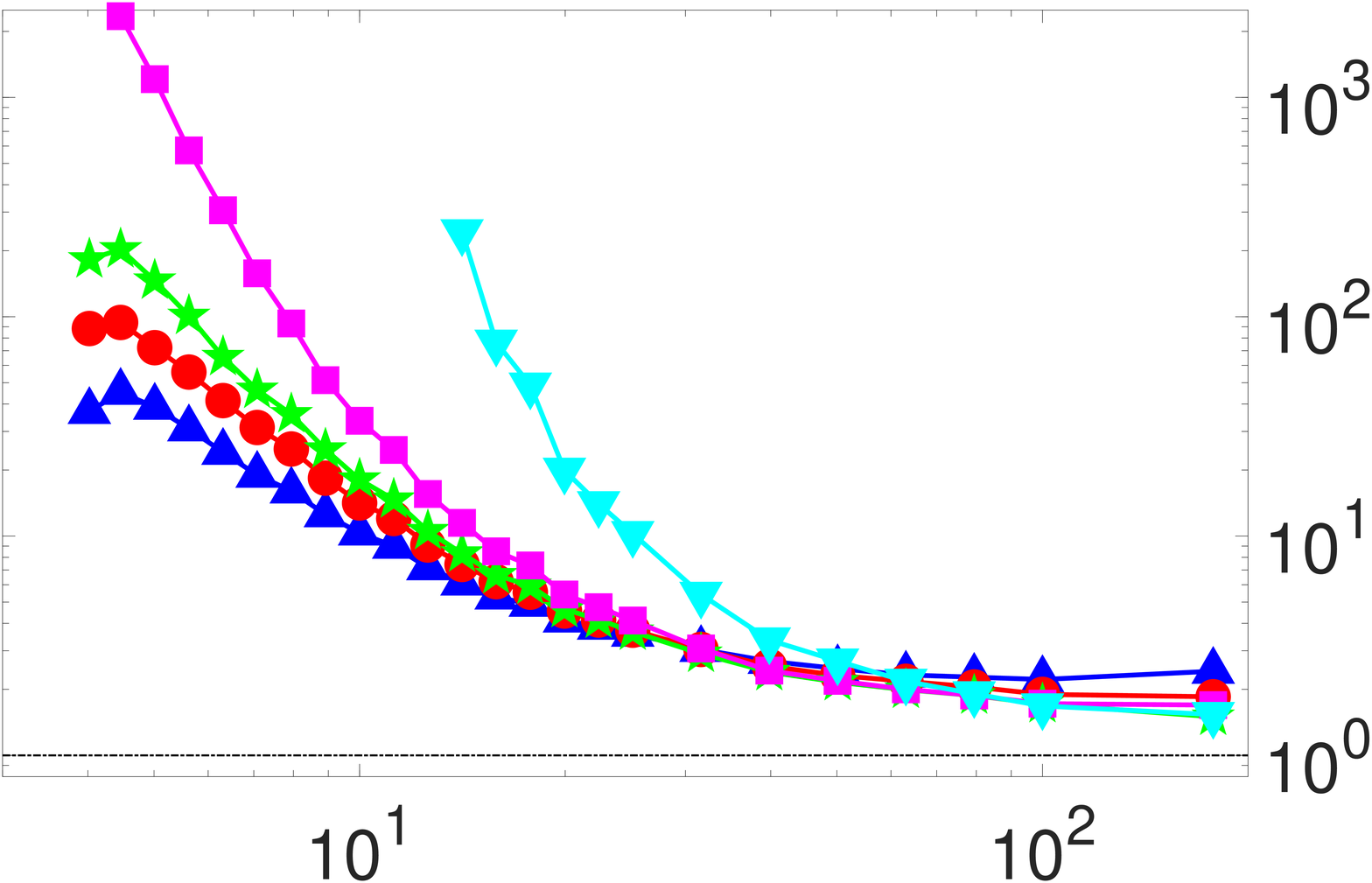}
\put(103,15){\rotatebox{90}{$ \frac{ \langle u'_{z,2WC}(\bm{x}^p(t),t)\rangle } {\langle u'_{z,1WC}(\bm{x}^p(t),t)\rangle} $}}
\put(48,-3){\footnotesize{$\ell_F / \eta$}}

\end{overpic}
}
	\end{overpic}}
\caption{Ratio of 1WC sub-grid settling velocity enhancement to that of the 2WC case shown as a function of the normalized filtering length $\ell_F/\eta$ for varying $St$ and fixed $Re_{\lambda} = 87$ and $Fr = 1$. Inset shows the ratio of 2WC to 1WC case and helps to infer magnitude of enhancement by 2WC. Inset is in log-log scale and some values at low $\ell_F / \eta$ are not shown here because of `noise' in the data.}
\label{SGSVel_1WC2WC_ratio_StDep_ScaleDep}
\end{figure}

\begin{figure}
\centering
\vspace{0mm}			
	{\begin{overpic}
	[trim = 0mm -25mm -55mm -3mm,
	scale=0.2,clip,tics=20]{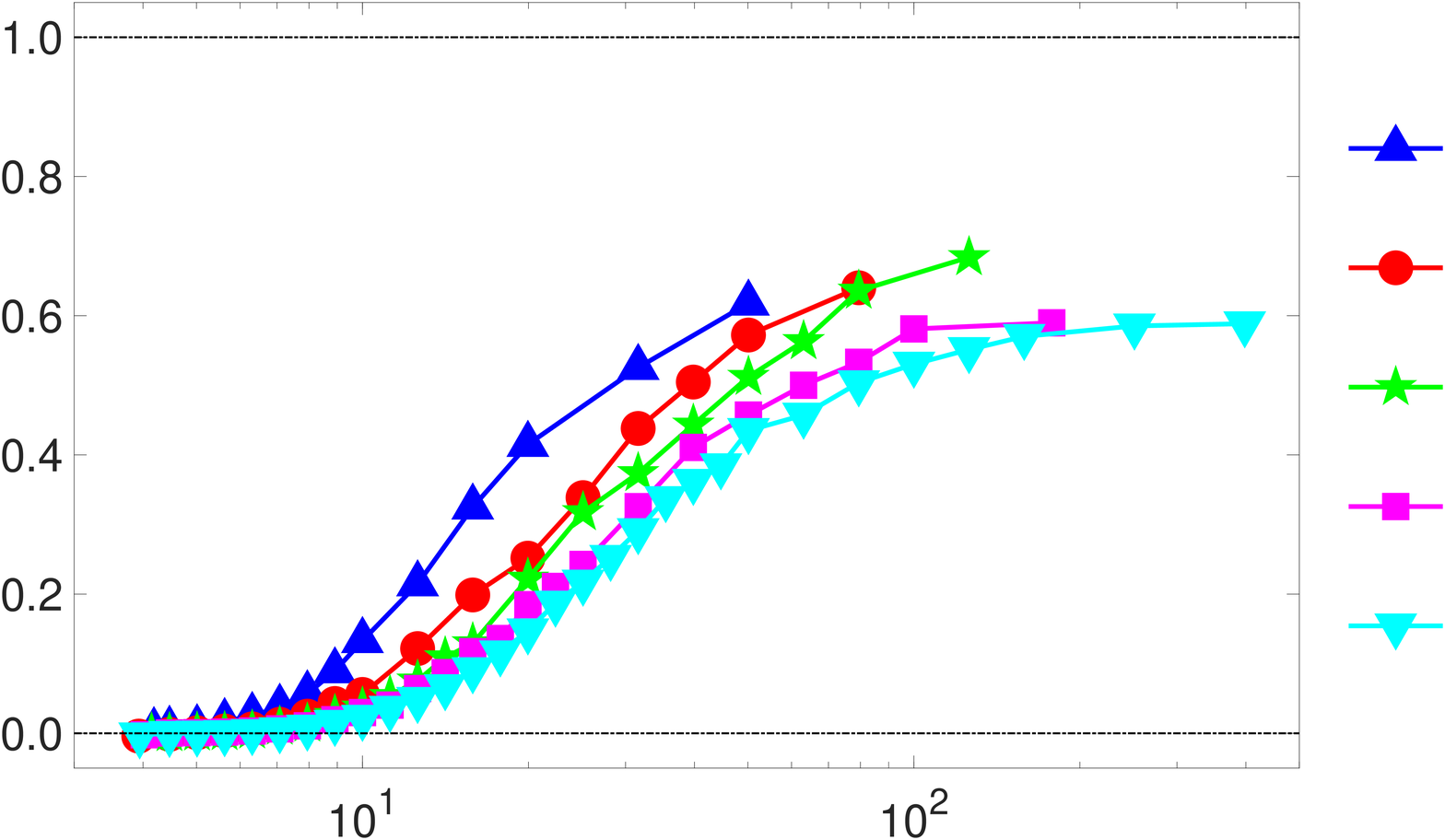}
\put(45,1){\large{$\ell_F / \eta$}}
\put(-10,22){\rotatebox{90}{\Large{$ \frac{ \langle u'_{z,1WC}(\bm{x}^p(t),t)\rangle } {\langle u'_{z,2WC}(\bm{x}^p(t),t)\rangle} $}}}
\put(92,46.75){\large{$Re_{\lambda} = 31$}}
\put(92,39.25){\large{$Re_{\lambda} = 54$}}
\put(92,31.75){\large{$Re_{\lambda} = 75$}}
\put(92,24.25){\large{$Re_{\lambda} = 87$}}
\put(92,16.75){\large{$Re_{\lambda} = 142$}}
\put(5,33){
\begin{overpic}[trim = 0mm 0mm 0mm 0mm, scale=0.075,clip,tics=20]
{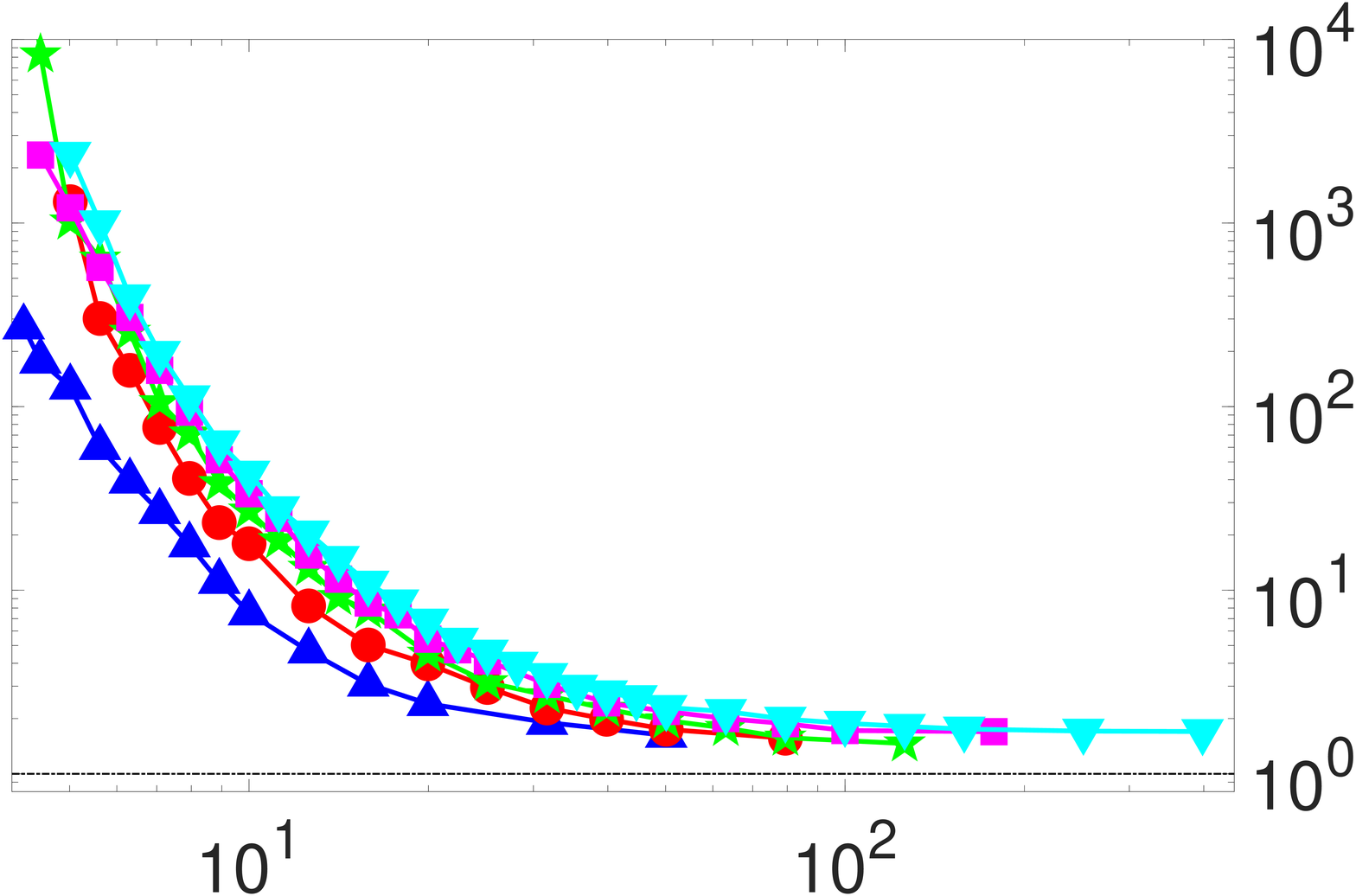}
\put(103,7.5){\rotatebox{90}{\small{$ \frac{ \langle u'_{z,2WC}(\bm{x}^p(t),t)\rangle } {\langle u'_{z,1WC}(\bm{x}^p(t),t)\rangle} $}}}
\put(35,-5){\footnotesize{$\ell_F / \eta$}}

\end{overpic}
}
	\end{overpic}}
\caption{Ratio of 1WC sub-grid settling velocity enhancement to that of the 2WC case shown as a function of the normalized filtering length $\ell_F/\eta$ for varying $Re_{\lambda}$ and fixed $St = 1$ and $Fr = 1$. Inset shows the ratio of 2WC to 1WC case and helps to infer magnitude of enhancement by 2WC. Inset is in log-log scale and some values at low $\ell_F / \eta$ are not shown here because of `noise' in the data.}
\label{SGSVel_1WC2WC_ratio_ReDep_ScaleDep}
\end{figure}


\begin{figure}
\centering
\vspace{0mm}			
	{\begin{overpic}
	[trim = 0mm -25mm -85mm -3mm,
	scale=0.19,clip,tics=20]{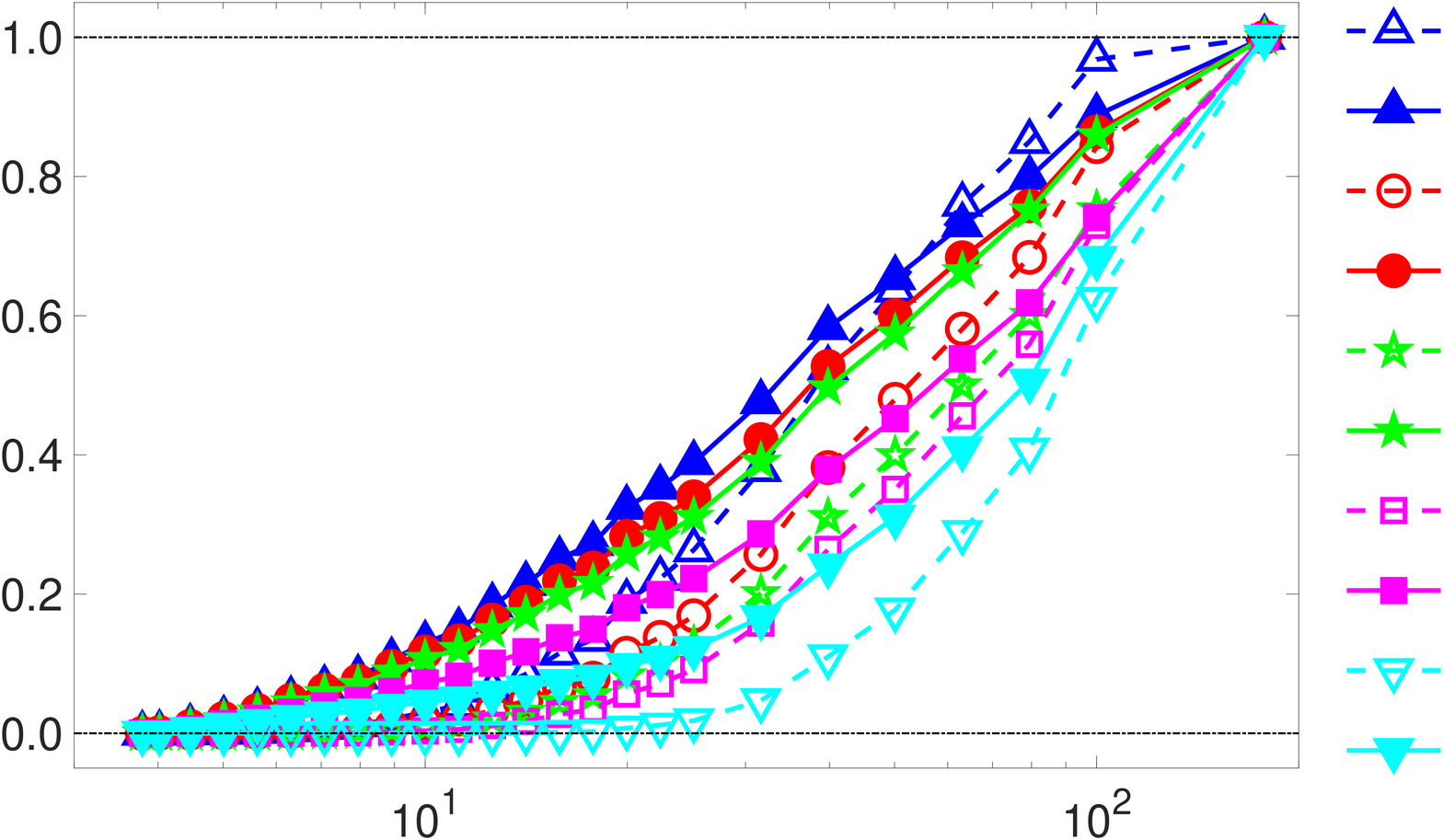}
\put(45,1){\large{$\ell_F / \eta$}}
\put(-9,22){\rotatebox{90}{\Large{$ \frac{ \langle u'_{z}(\bm{x}^p(t),t)\rangle } {\langle u_{z}(\bm{x}^p(t),t)\rangle} $}}}
\put(87.5,51.55){\large{$St = 0.3$ 1WC}}
\put(87.5,46.75){\large{$St = 0.3$ 2WC}}
\put(87.5,41.95){\large{$St = 0.5$ 1WC}}
\put(87.5,37.15){\large{$St = 0.5$ 2WC}}
\put(87.5,32.35){\large{$St = 0.7$ 1WC}}
\put(87.5,27.55){\large{$St = 0.7$ 2WC}}
\put(87.5,22.75){\large{$St = 1.0$ 1WC}}
\put(87.5,17.95){\large{$St = 1.0$ 2WC}}
\put(87.5,13.15){\large{$St = 2.0$ 1WC}}
\put(87.5,8.35){\large{$St = 2.0$ 2WC}}
	\end{overpic}}
\caption{Ratio of sub-grid settling velocity enhancement to the `full' settling velocity enhancement for the 1WC (open symbols, dashed line) and 2WC  (closed symbols, solid line) cases shown as a function of the normalized filtering length $\ell_F/\eta$ for varying $St$ and fixed $Re_{\lambda} = 87$ and $Fr = 1$.}
\label{SGS_by_Unfilt_Vel_StDep_ScaleDep}
\end{figure}

\begin{figure}
\centering
\vspace{0mm}			
	{\begin{overpic}
	[trim = 0mm -25mm -85mm -3mm,
	scale=0.19,clip,tics=20]{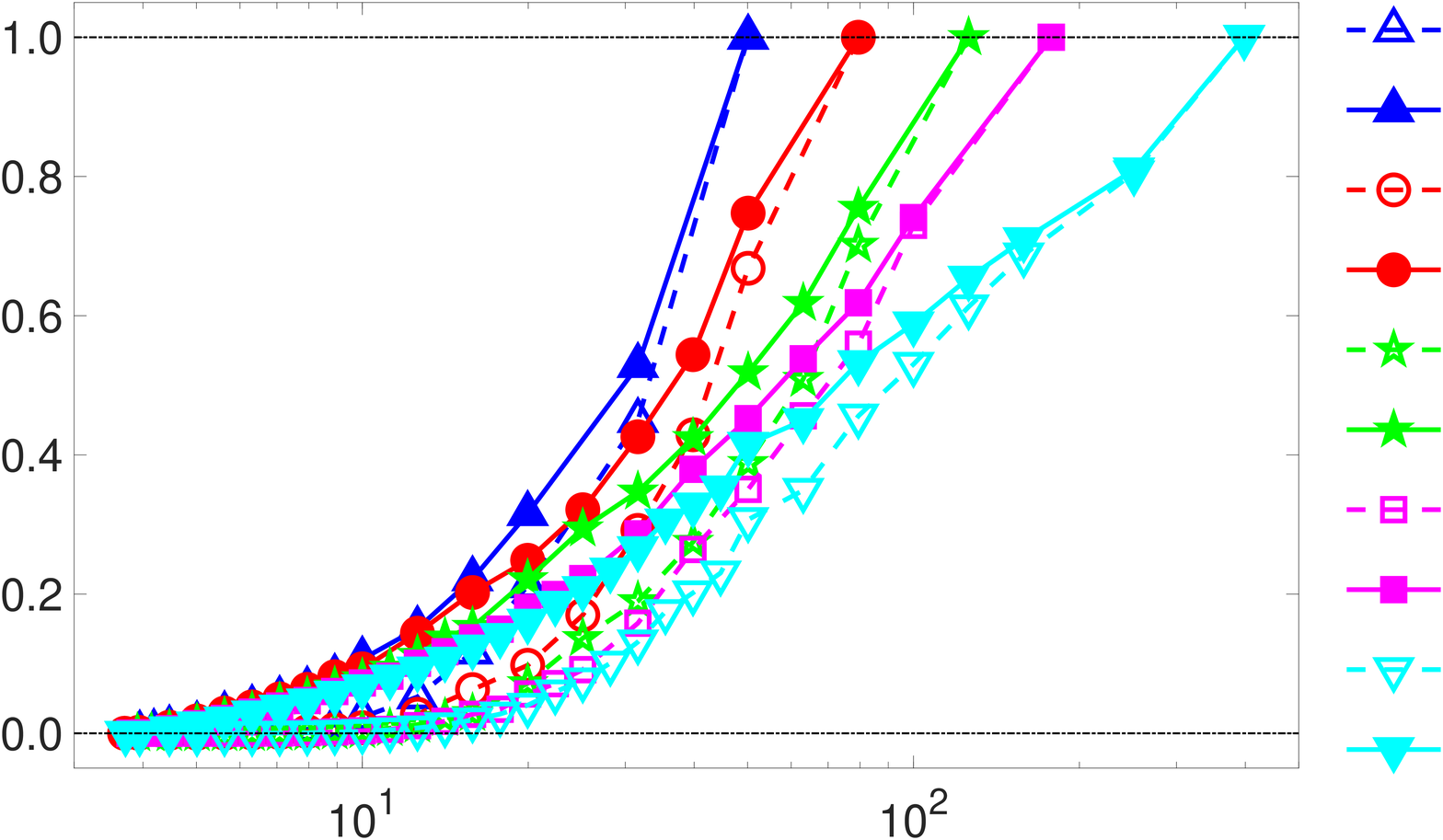}
\put(45,1){\large{$\ell_F / \eta$}}
\put(-9,22){\rotatebox{90}{\Large{$ \frac{ \langle u'_{z}(\bm{x}^p(t),t)\rangle } {\langle u_{z}(\bm{x}^p(t),t)\rangle} $}}}
\put(87.5,51.55){\large{$Re_{\lambda} = 31$ 1WC}}
\put(87.5,46.75){\large{$Re_{\lambda} = 31$ 2WC}}
\put(87.5,41.95){\large{$Re_{\lambda} = 54$ 1WC}}
\put(87.5,37.15){\large{$Re_{\lambda} = 54$ 2WC}}
\put(87.5,32.35){\large{$Re_{\lambda} = 75$ 1WC}}
\put(87.5,27.55){\large{$Re_{\lambda} = 75$ 2WC}}
\put(87.5,22.75){\large{$Re_{\lambda} = 87$ 1WC}}
\put(87.5,17.95){\large{$Re_{\lambda} = 87$ 2WC}}
\put(87.5,13.15){\large{$Re_{\lambda} = 142$ 1WC}}
\put(87.5,8.35){\large{$Re_{\lambda} = 142$ 2WC}}
	\end{overpic}}
\caption{Ratio of sub-grid settling velocity enhancement to the `full' settling velocity enhancement for the 1WC (open symbols, dashed line) and 2WC (closed symbols, solid line) cases shown as a function of the normalized filtering length $\ell_F/\eta$ for varying $Re_{\lambda}$ and fixed $St = 1$ and $Fr = 1$.}
\label{SGS_by_Unfilt_Vel_ReDep_ScaleDep}
\end{figure}


\subsection{Contribution of different scales to settling enhancement}

Our previous study \citet{tom19} highlighted the role that different flows scales have in enhancing the particle settling velocities via the multiscale preferential sweeping mechanism. We now explore how 2WC influences the role of these different scales in contributing to the particle settling velocity enhancement.

In figures \ref{SGSVel_1WC2WC_ratio_StDep_ScaleDep} and \ref{SGSVel_1WC2WC_ratio_ReDep_ScaleDep}, we consider the ratio of $\langle{u'}_z(\bm{x}^p(t),t)\rangle$ for the 1WC and 2WC flows for varying $St$ and $Re_{\lambda}$, respectively. Here, $u_z'$ denotes the vertical fluid velocity field with scales $\geq\ell_F$ filtered out (i.e.~it is the `sub-grid field'), such that $\langle{u'}_z(\bm{x}^p(t),t)\rangle$ denotes the contribution of all scales less than $\ell_F$ to the full quantity $\langle{u}_z(\bm{x}^p(t),t)\rangle$. The results show a very strong modification of $\langle{u'}_z(\bm{x}^p(t),t)\rangle$ due to 2WC at the small scales. Indeed, the insets in figures \ref{SGSVel_1WC2WC_ratio_StDep_ScaleDep} and \ref{SGSVel_1WC2WC_ratio_ReDep_ScaleDep} show that 2WC enhances $\langle{u'}_z(\bm{x}^p(t),t)\rangle$ at small scales by 2 to 3 orders of magnitude. As $\ell_F$ is increased, the impact of 2WC on $\langle{u'}_z(\bm{x}^p(t),t)\rangle$ reduces. However, its effect is still significant at the largest $\ell_F$ considered for which $\langle{u'}_z(\bm{x}^p(t),t)\rangle=\langle{u}_z(\bm{x}^p(t),t)\rangle$ showing that $\langle{u}_z(\bm{x}^p(t),t)\rangle$ can be more than twice as large for the 2WC case compared to the 1WC case. For low values of $\ell_F$, say $\ell_F/\eta < 20$, we see that the effect of 2WC is larger for higher $St$. Hence, at these small scales, increasing $St$ increases the impact of 2WC on $\langle{u'}_z(\bm{x}^p(t),t)\rangle$. In a similar range of scales, increasing $Re_{\lambda}$ for a fixed $St$ increases the influence of 2WC on $\langle{u'}_z(\bm{x}^p(t),t)\rangle$. These results also show that the effect of 2WC is important at all scales in the flow since $|\langle{u'}_z(\bm{x}^p(t),t)\rangle|$ is greater for the 2WC than the 1WC case at all scales considered. The fact that for the larger for larger $Re_{\lambda} $ cases the ratio of $\langle{u'}_z(\bm{x}^p(t),t)\rangle$ for the 2WC to the 1WC case approaches a constant value as $\ell_F$ is increased is probably simply because there $\ell_F$ is approaching the largest scales in the flow. We will return later to consider the physical mechanisms responsible for 2WC leading to a strong enhancement of $\langle{u}_z(\bm{x}^p(t),t)\rangle$.

As just highlighted, although the effect of 2WC on $\langle{u'}_z(\bm{x}^p(t),t)\rangle$ is very strong at the smallest scales, its effect on the full quantity $\langle{u}_z(\bm{x}^p(t),t)\rangle$ is less dramatic. To understand why this is so we look at the contribution of different scales to the full quantity $\langle{u}_z(\bm{x}^p(t),t)\rangle$ in figures \ref{SGS_by_Unfilt_Vel_StDep_ScaleDep} and \ref{SGS_by_Unfilt_Vel_ReDep_ScaleDep}. We see that the smallest scales where the effect of 2WC is most significant, i.e.~$\ell_F / \eta \lesssim 20$, only contribute about $10 - 20 \%$ of the full quantity $\langle{u}_z(\bm{x}^p(t),t)\rangle$. Note that in these results, $\langle{u'}_z(\bm{x}^p(t),t)\rangle/\langle{u}_z(\bm{x}^p(t),t)\rangle$ does not saturate with increasing $\ell_F/\eta$, but only reaches a value of unity when all scales of the flow are included. According to the theory in \cite{tom19}, for $Re_\lambda\to\infty$ the quantity $\langle{u'}_z(\bm{x}^p(t),t)\rangle/\langle{u}_z(\bm{x}^p(t),t)\rangle$ would approach unity in the limit $\ell_F\to\ell_c(St)\ll L$ (assuming $St$ is finite), where $\ell_c(St)$ is defined as the scale above which the effects of particle inertia are negligible. Since 2WC arises due to the particles not following the local fluid velocity field due to their inertia, then $\ell_c(St)$ must also provide an upper bound on the scale at which 2WC can impact the flow. The $Re_\lambda$ of our DNS are too small to observe these asymptotic regimes which are, however, likely be of importance in atmospheric contexts where typically $Re_\lambda\geq (10^4)$.

\subsection{Effectiveness of the preferential sweeping mechanism in 2WC flows}\label{PrefSweep2WC}

Having explored the role of different flow scales that contribute to $\langle u_z(\bm{x}^p(t),t)\rangle$, we now turn to consider the role played by the preferential sweeping mechanism in governing $\langle u_z(\bm{x}^p(t),t)\rangle$, and to understand how 2WC leads to the strong enhancement of this quantity that was observed earlier. This is especially important to consider since \cite{monchaux17} argued that in the presence of 2WC, the role of preferential sweeping is reduced, and possibly even eradicated, and that instead $\langle u_z(\bm{x}^p(t),t)\rangle<0$ is mainly due to the particles dragging the surrounding fluid down with them as they fall. 

Many previous studies have sought to detect the role of preferential sweeping by considering the fluid statistics conditioned on the local particle concentration. However, as discussed in \cite{tom19}, this is inappropriate for investigating the role of the preferential sweeping mechanism. Instead, one needs to consider the local fluid velocity conditioned on the local flow structure (since the preferential sweeping mechanism specifically argues that the settling enhancement is due to the particle motion in strain dominated regions of the flow), and to consider this we introduce the invariant
\begin{align}
{\mathcal{Q}}^p(t)\equiv {\mathcal{S}}^2(\bm{x}^p(t),t)-{\mathcal{R}}^2(\bm{x}^p(t),t),
\end{align}
with which we may write
\begin{align}
\langle u_z(\bm{x}^p(t),t)\rangle=\int_{\mathbb{R}}\Big\langle u_z(\bm{x}^p(t),t)\Big\rangle_{{\mathcal{Q}}}\,\mathcal{P}({\mathcal{Q}},t)\, d{\mathcal{Q}},
\end{align}
where $\mathcal{P}({\mathcal{Q}},t)\equiv \langle \delta({\mathcal{Q}}^p(t)-{\mathcal{Q}})\rangle$ is the probability density function of ${\mathcal{Q}}^p(t)$, and $\mathcal{Q}$ is the corresponding sample-space coordinate. Here, $\mathcal{S}^2$ and $\mathcal{R}^2$ are the second invariants of the strain-rate $\bm{\mathcal{S}}\equiv(\bm{\nabla u}+\bm{\nabla u}^\top)/2$ and rotation-rate $\bm{\mathcal{R}}\equiv(\bm{\nabla u}-\bm{\nabla u}^\top)/2$ tensors, respectively.

We may then define 

\begin{align}
A(\alpha)&\equiv\int_{-\infty}^{-\alpha}\Big\langle u_z(\bm{x}^p(t),t)\Big\rangle_{{\mathcal{Q}}}\,\mathcal{P}({\mathcal{Q}},t)\, d{\mathcal{Q}},\\
B(\alpha)&\equiv\int_{\alpha}^{\infty}\Big\langle u_z(\bm{x}^p(t),t)\Big\rangle_{{\mathcal{Q}}}\,\mathcal{P}({\mathcal{Q}},t)\, d{\mathcal{Q}}.
\end{align}
According to the preferential sweeping mechanism, we would expect $|B(0)|>|A(0)|$ with $B(0)<0$, i.e.~ $\langle u_z(\bm{x}^p(t),t)\rangle$ is dominated by contributions from particles in strain dominated regions of the flow (where ${\mathcal{Q}}>0$), and in these regions the particles experience on average a downward moving fluid velocity.

\begin{figure}
\centering
\vspace{0mm}			
	{\begin{overpic}
	[trim = 0mm -25mm -95mm -3mm,
	scale=0.175,clip,tics=20]{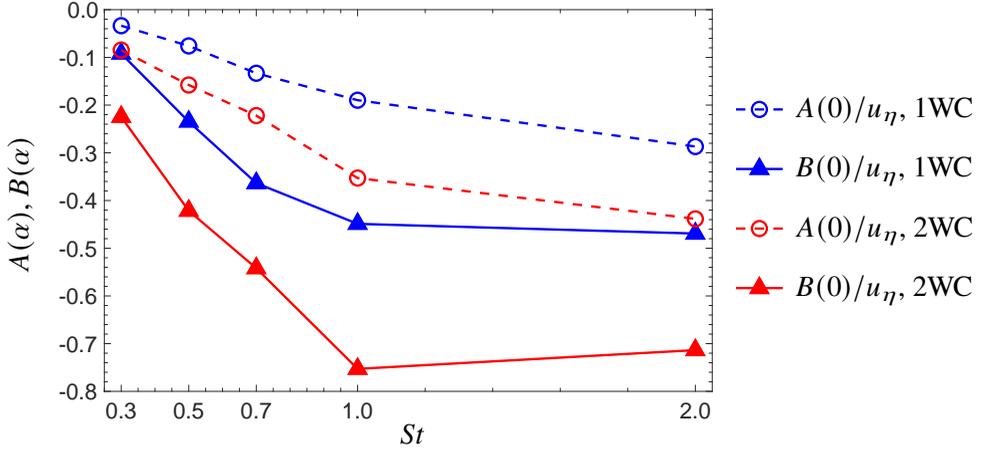}
	\put(-6,21){\rotatebox{90}{\large{$A(\alpha),B(\alpha)$}}}
	\put(40,1){\large{$St$}}		
	\put(86.5,39.25){\large{$A(0)/u_\eta$, 1WC}}
	\put(86.5,32.25){\large{$B(0)/u_\eta$, 1WC}}		
	\put(86.5,25.25){\large{$A(0)/u_\eta$, 2WC}}
	\put(86.5,18.25){\large{$B(0)/u_\eta$, 2WC}}
	\end{overpic}}
\caption{DNS results for $A(0)/u_\eta$ and $B(0)/u_\eta$ as a function of $St$ and for the 1WC and 2WC cases. Here, $A(0)$ denotes the total contribution to $\langle u_z(\bm{x}^p(t),t)\rangle$ from particles in rotation dominated regions of the flow, while $B(0)$ denotes the total contribution to $\langle u_z(\bm{x}^p(t),t)\rangle$ from particles in strain dominated regions of the flow.}
\label{u_av_Q}
\end{figure}

\begin{figure}
\centering
\vspace{0mm}			
	{\begin{overpic}
	[trim = 0mm -25mm -45mm -3mm,
	scale=0.2,clip,tics=20]{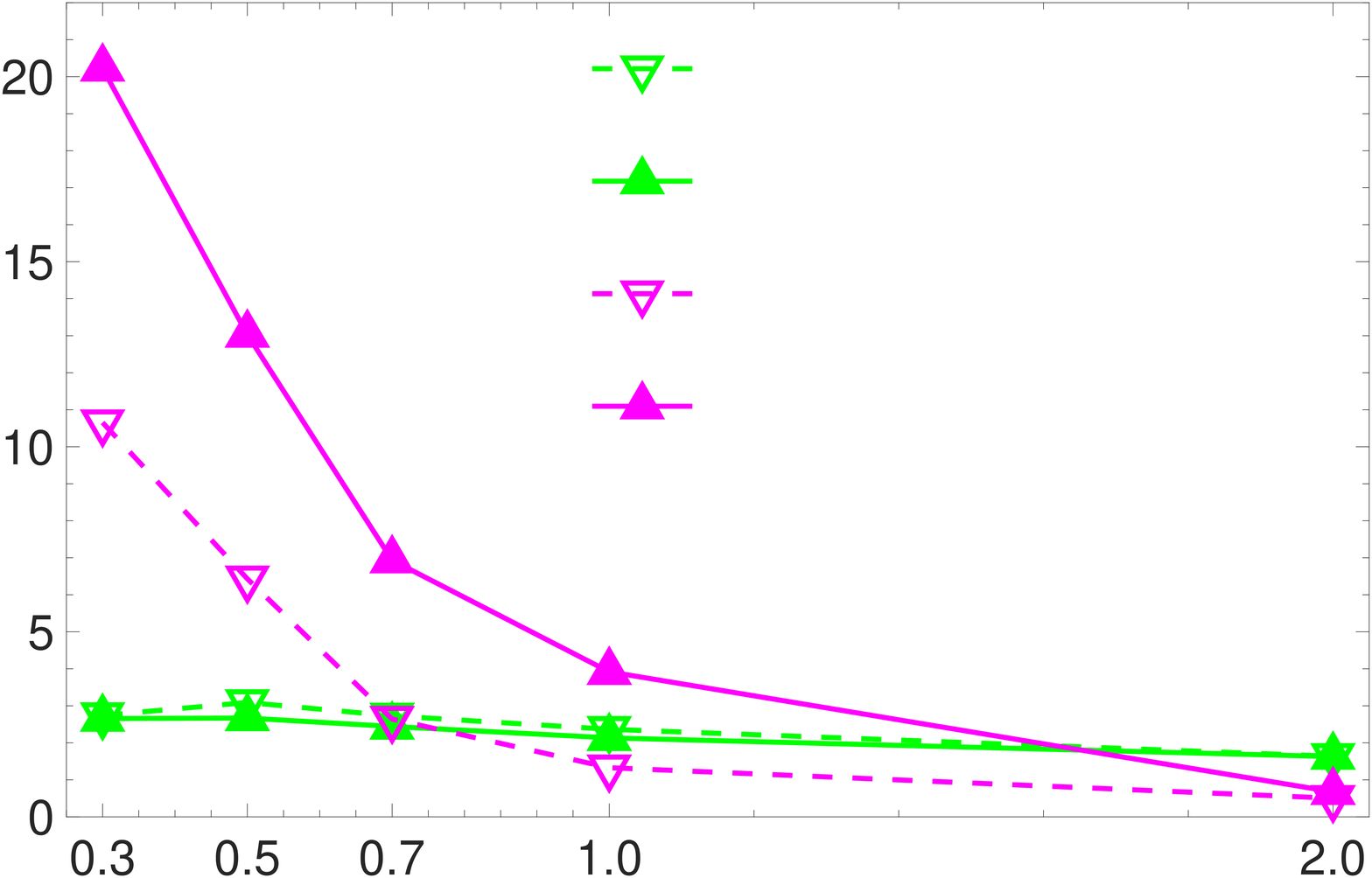}
	\put(-6,25){\rotatebox{90}{\large{$A(\alpha)/B(\alpha)$}}}
	\put(45,1){\large{$St$}}			
	\put(48,58){\large{$B(\alpha_1)/A(\alpha_1)$, 1WC}}
	\put(48,50.5){\large{$B(\alpha_1)/A(\alpha_1)$, 2WC}}			
	\put(48,43){\large{$B(\alpha_2)/A(\alpha_2)$, 1WC}}
	\put(48,35.5){\large{$B(\alpha_2)/A(\alpha_2)$, 2WC}}	
	\end{overpic}}
\caption{DNS results for $B(\alpha)/A(\alpha)$ as a function of $St$ and for the 1WC and 2WC cases. In the plots, $\alpha_1=0$ and $\alpha_2=4\langle\mathcal{S}^2\rangle$. Here, $A(\alpha)$ denotes the total contribution to $\langle u_z(\bm{x}^p(t),t)\rangle$ from particles in regions of the flow where $\mathcal{Q}<-\alpha$, while $B(\alpha)$ denotes the total contribution to $\langle u_z(\bm{x}^p(t),t)\rangle$ from particles in regions of the flow where $\mathcal{Q}>\alpha$.}
\label{u_av_Q_ratio}
\end{figure}
\FloatBarrier
The results in figure \ref{u_av_Q} clearly show that $|B(0)|>|A(0)|$ and $B(0)<0$ for both the 1WC and 2WC cases, confirming the role of the preferential sweeping mechanism in both cases.  It is interesting to note that even for the 1WC case $A(0)$ is also negative, implying that even when the particles are in a rotation dominated region, they still experience an enhanced settling due to the turbulence. While this may seem surprising, it is not inconsistent with the preferential sweeping mechanism for at least two reasons. First, the preferential sweeping mechanism simply argues that the dominant contribution to $\langle u_z(\bm{x}^p(t),t)\rangle$ is from particles that move in strain dominated regions, and does not say anything explicitly about what the contribution to $\langle u_z(\bm{x}^p(t),t)\rangle$ is from the sub-set of particles that move through rotation dominated regions of the flow. Second, if the inertial particle falls through a vortex (where $\mathcal{Q}<0$), then provided it still has the tendency to fall through the downward side of the vortex, then in this region the particle would also experience a downward moving flow, enabling one to have $\langle u_z(\bm{x}^p(t),t)\rangle_{\mathcal{Q}<0}<0$.

The results in figure \ref{u_av_Q} show that 2WC enhances both $|A(0)|$ and $|B(0)|$, and this is plausibly explained in terms of the fluid dragging effect discussed in \cite{monchaux17}. Namely, as the particles fall through the flow, they tend to accelerate the fluid downward on average, which then enhances the downward velocity of the fluid in the vicinity of the particles. Nevertheless, the results in figure \ref{u_av_Q} imply that this dragging effect occurs in both strain and rotation dominated regions of the flow, and that overall the turbulence enhancement of the particle settling velocity continues to be dominated by contributions from particles in strain dominated regions of the flow. Put together, this means that $\langle u_z(\bm{x}^p(t),t)\rangle$ is dominated by contributions from particles in strain dominated regions of the flow, and that for the 2WC case, in these regions the particles also drag the fluid with them, increasing the local fluid velocity and hence enhancing $\langle u_z(\bm{x}^p(t),t)\rangle$ compared to the 1WC case. These results then provide strong evidence that the preferential sweeping mechanism continues to be the basic mechanism by which the turbulence enhances the particle settling velocity compared with the Stokes settling velocity.

In figure \ref{u_av_Q_ratio} we show the ratios $B(\alpha_1)/A(\alpha_1)$ and $B(\alpha_2)/A(\alpha_2)$ with $\alpha_1=0$ and $\alpha_2=4\langle\mathcal{S}^2\rangle$. The results show that $B(\alpha_1)/A(\alpha_1)$ is greater than 1 and that 2WC has a relatively small effect on this ratio. On the other hand, the results for $B(\alpha_2)/A(\alpha_2)$ are much larger and show a much stronger effect of 2WC. These results with $\alpha_2=4\langle\mathcal{S}^2\rangle$ only consider contributions from regions of strong strain ($\mathcal{Q}>4\langle\mathcal{S}^2\rangle$) and strong rotation ($\mathcal{Q}<-4\langle\mathcal{S}^2\rangle$), and isolating these regions reveals a very strong effect of preferential sweeping. This is most likely due to the fact that $A$ and $B$ for $\alpha=0$ are dominated by contributions to their integrals from regions where $\mathcal{Q}$ is not that large, e.g.  $|\mathcal{Q}|\leq O(\langle \mathcal{S}^2\rangle)$, and in such a region, $\mathcal{Q}<0$ might be associated with rotational fluid motion that is also accompanied by significant straining, i.e.~not regions of solid-body rotation. On the other hand, regions where $\mathcal{Q}\ll -\langle \mathcal{S}^2\rangle$ are more likely to be associated with regions of quasi-solid-body rotation, and it is these regions that are most effective in centrifuging the inertial particles.

\begin{figure}
\centering
\vspace{0mm}			
{\begin{overpic}[trim = 0mm -25mm -45mm -3mm,
scale=0.2,clip,tics=20]{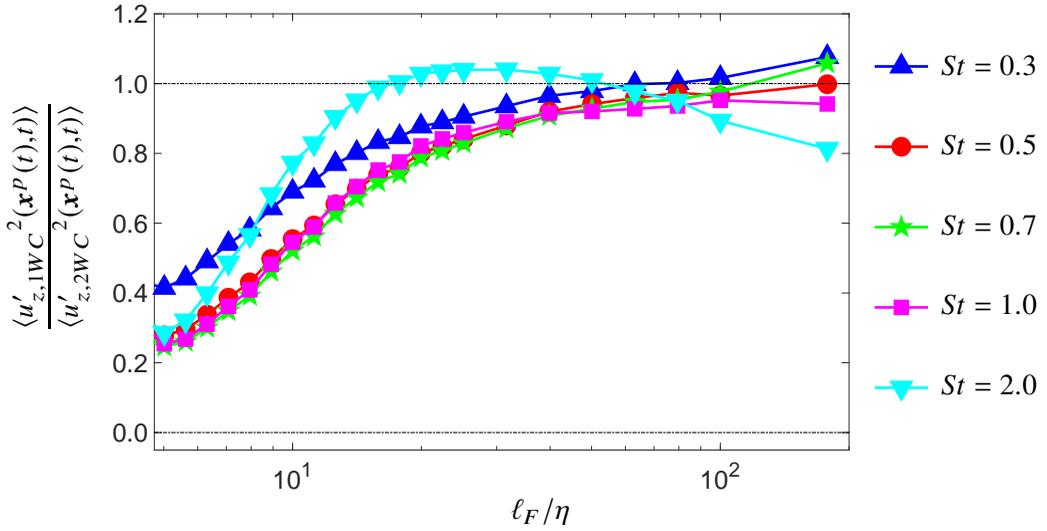}
\put(45,1){\large{$\ell_F / \eta$}}
\put(-12,22){\rotatebox{90}{\Large{$ \frac{ \langle {u'_{z,1WC}}^2(\bm{x}^p(t),t)\rangle } {\langle {u'_{z,2WC}}^2(\bm{x}^p(t),t)\rangle} $}}}
\put(93.5,51){\large{$St = 0.3$}}
\put(93.5,42){\large{$St = 0.5$}}
\put(93.5,33){\large{$St = 0.7$}}
\put(93.5,24){\large{$St = 1.0$}}
\put(93.5,15){\large{$St = 2.0$}}
\end{overpic}}
\caption{Ratio of variances of 1WC sub-grid settling velocity enhancement to that of the 2WC case shown as a function of the normalized filtering length $\ell_F/\eta$ for varying $St$ and fixed $Re_{\lambda} = 87$ and $Fr = 1$.}
\label{Var_SGSVel_StDep_ScaleDep}
\end{figure}

\begin{figure}
\centering
\vspace{0mm}			
	{\begin{overpic}
	[trim = 0mm -25mm -45mm -3mm,
	scale=0.2,clip,tics=20]{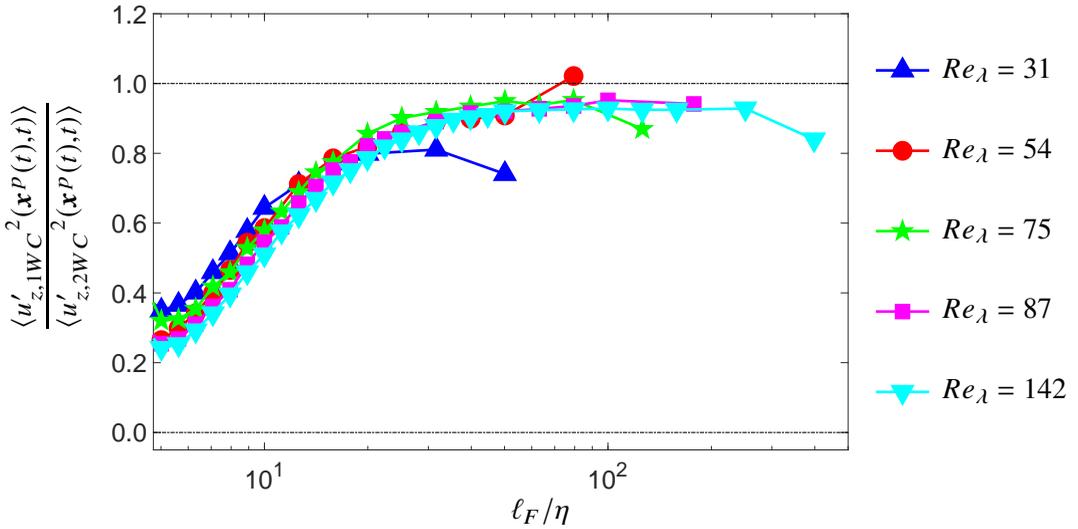}
\put(45,1){\large{$\ell_F / \eta$}}
\put(-12,22){\rotatebox{90}{\Large{$ \frac{ \langle {u'_{z,1WC}}^2(\bm{x}^p(t),t)\rangle } {\langle {u'_{z,2WC}}^2(\bm{x}^p(t),t)\rangle} $}}}
\put(93.5,50.5){\large{$Re_{\lambda} = 31$}}
\put(93.5,41.5){\large{$Re_{\lambda} = 54$}}
\put(93.5,32.5){\large{$Re_{\lambda} = 75$}}
\put(93.5,23.5){\large{$Re_{\lambda} = 87$}}
\put(93.5,14.5){\large{$Re_{\lambda} = 142$}}
	\end{overpic}}
\caption{Ratio of variances of 1WC sub-grid settling velocity enhancement to that of the 2WC case shown as a function of the normalized filtering length $\ell_F/\eta$ for varying $Re_{\lambda}$ and fixed $St = 1$ and $Fr = 1$.}
\label{Var_SGSVel_ReDep_ScaleDep}
\end{figure}

Since $\langle u'_z(\bm{x}^p(t),t)\rangle$ is an odd moment that depends delicately on how the settling particles sample the flow, another way to substantiate the idea that 2WC enhances the particle settling compared with the 1WC due to the fluid dragging effect is to also consider the influence of 2WC on the second moment $\langle u'^{2}_z(\bm{x}^p(t),t)\rangle$ which would be finite even for non-settling, non-inertial particles. In figures \ref{Var_SGSVel_StDep_ScaleDep} and \ref{Var_SGSVel_ReDep_ScaleDep}, we show the ratio of $\langle u'^{2}_z(\bm{x}^p(t),t)\rangle$ for the 1WC to the 2WC. It can be clearly seen that the fluid velocity fluctuations sampled by the particle are larger in the 2WC case, which is again plausibly due to the fluid dragging effect discussed in \cite{monchaux17}. Comparing the results in figures \ref{Var_SGSVel_StDep_ScaleDep} and \ref{Var_SGSVel_ReDep_ScaleDep} to those in figures \ref{SGSVel_1WC2WC_ratio_StDep_ScaleDep} and \ref{SGSVel_1WC2WC_ratio_ReDep_ScaleDep} shows that the first moment of $u'_z(\bm{x}^p(t),t)$ is more sensitive to 2WC than the second moments. This is likely due to the more general fact that odd moments of flow quantities evaluated along particle trajectories are often much more sensitive to the effects of particle inertia (and therefore 2WC) than even moments since odd moments depend sensitively on cancellation between positive and negative values of the random variable (e.g. see figure 6 in \cite{bragg17b}).


\FloatBarrier

In contrast to \cite{monchaux17} therefore, these results provide strong evidence that, at least for the $\Phi_m$ considered, preferential sweeping continues to be the mechanism responsible for the enhanced particle settling, with the contribution of 2WC being mainly that in the downward, strain dominated regions where the particles preferentially move due to the centrifuging effect, the particles drag the fluid down with them, further enhancing their settling compared to that produced by preferential sweeping in the 1WC case. The result that led \cite{monchaux17} to conclude that 2WC strongly reduces the role of preferential sweeping was based on their observation that if they compute the PDF of the fluid velocity gradients along the settling inertial particle trajectories, then as either $Sv$ or $\Phi_m$ is increased, the preferential sampling of the flow (on which the preferential sweeping mechanism depends) is dramatically reduced. However, contrary to their interpretation, this does not necessarily mean that the preferential sweeping mechanism is ceasing to play a role in governing the particle settling. Indeed, as was explained and demonstrated in \cite{tom19}, even for a 1WC system, as $Sv$ is increased, the scales of the flow at which the preferential sweeping mechanism operates shifts to larger scales. As such, for sufficiently large $Sv$, one will not observe preferential sampling of the unfiltered velocity gradients (as was observed in \cite{monchaux17}), but one may nevertheless observe preferential sampling of the velocity gradients filtered at a suitable scale. 

\begin{figure}
\centering
\vspace{0mm}			
{\begin{overpic}
[trim = 0mm -25mm -95mm -3mm, 
scale=0.2,clip,tics=20]{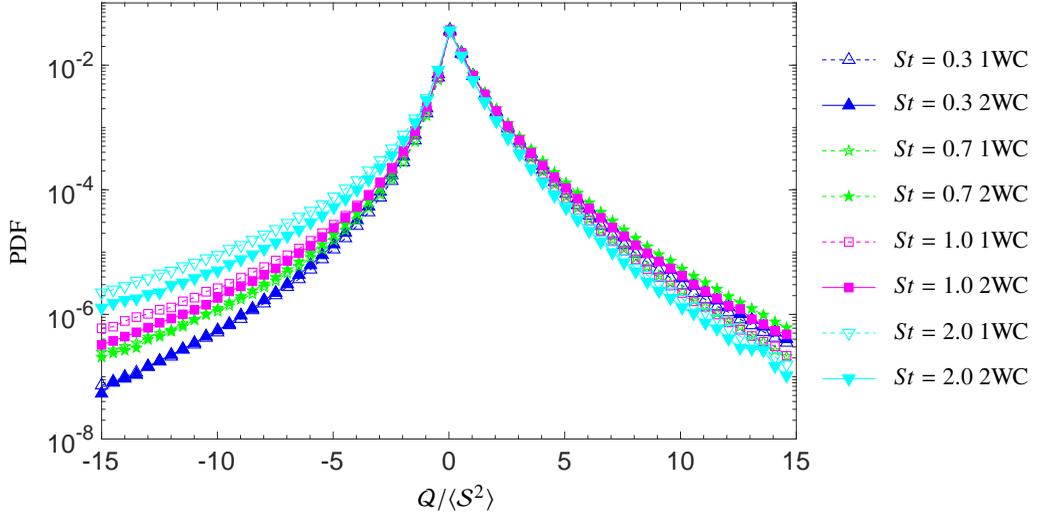}
\put(38,0.25){$\mathcal{Q}/\langle\mathcal{S}^2\rangle$}
\put(-4,25){\rotatebox{90}{PDF}}
\put(87,45.25){{$St = 0.3$ 1WC}}
\put(87,40.82){{$St = 0.3$ 2WC}}
\put(87,36.14){{$St = 0.7$ 1WC}}
\put(87,31.46){{$St = 0.7$ 2WC}}
\put(87,26.78){{$St = 1.0$ 1WC}}
\put(87,22.10){{$St = 1.0$ 2WC}}
\put(87,17.42){{$St = 2.0$ 1WC}}
\put(87,12.74){{$St = 2.0$ 2WC}}
\end{overpic}}
\caption{DNS results for $\mathcal{P}(\mathcal{Q})$ for different $St$, and for the 1WC and 2WC cases.}
\label{PDF_Q}
\end{figure}

With respect to how 2WC might further modify preferential sampling, above and beyond the effect of $Sv$ just discussed, in figure \ref{PDF_Q} we show plots of the PDF of $\mathcal{Q}^p(t)$ for different $St$, and for the 1WC and 2WC cases. Our results show that 2WC has a weak effect on these statistics, implying that the preferential sampling of the flow by the particles is only weakly affected by 2WC, at least for the $\Phi_m$ we have considered. Taken together with our previous point about the effect of $Sv$, this then implies that the strong reduction of preferential sampling of the flow observed by \cite{monchaux17} for their lower $\Phi_m$ case (which is the same $\Phi_m$ as we consider) is not due to 2WC, but is rather due to the effect of increasing $Sv$. Our results in figure \ref{u_av_Q} provide convincing evidence that preferential sweeping continues to be the mechanism enhancing the particle settling even in the presence of 2WC. For larger $Sv$ (corresponding to smaller $Fr$), we expect that this will still be the case, although the preferential sweeping will take place at larger scales in the flow (as was observed in the 1WC case in \cite{tom19}). This, as well as the question of how large $\Phi_m$ must become before the preferential sweeping mechanism ceases to be effective, are important questions for investigation in future work.

\FloatBarrier

\newpage\subsection{Effect of 2WC on preferential sampling at different scales}


\begin{figure}
\centering
\vspace{0mm}			
{\begin{overpic}
[trim = 0mm -25mm -82mm -3mm,
scale=0.20,clip,tics=20]{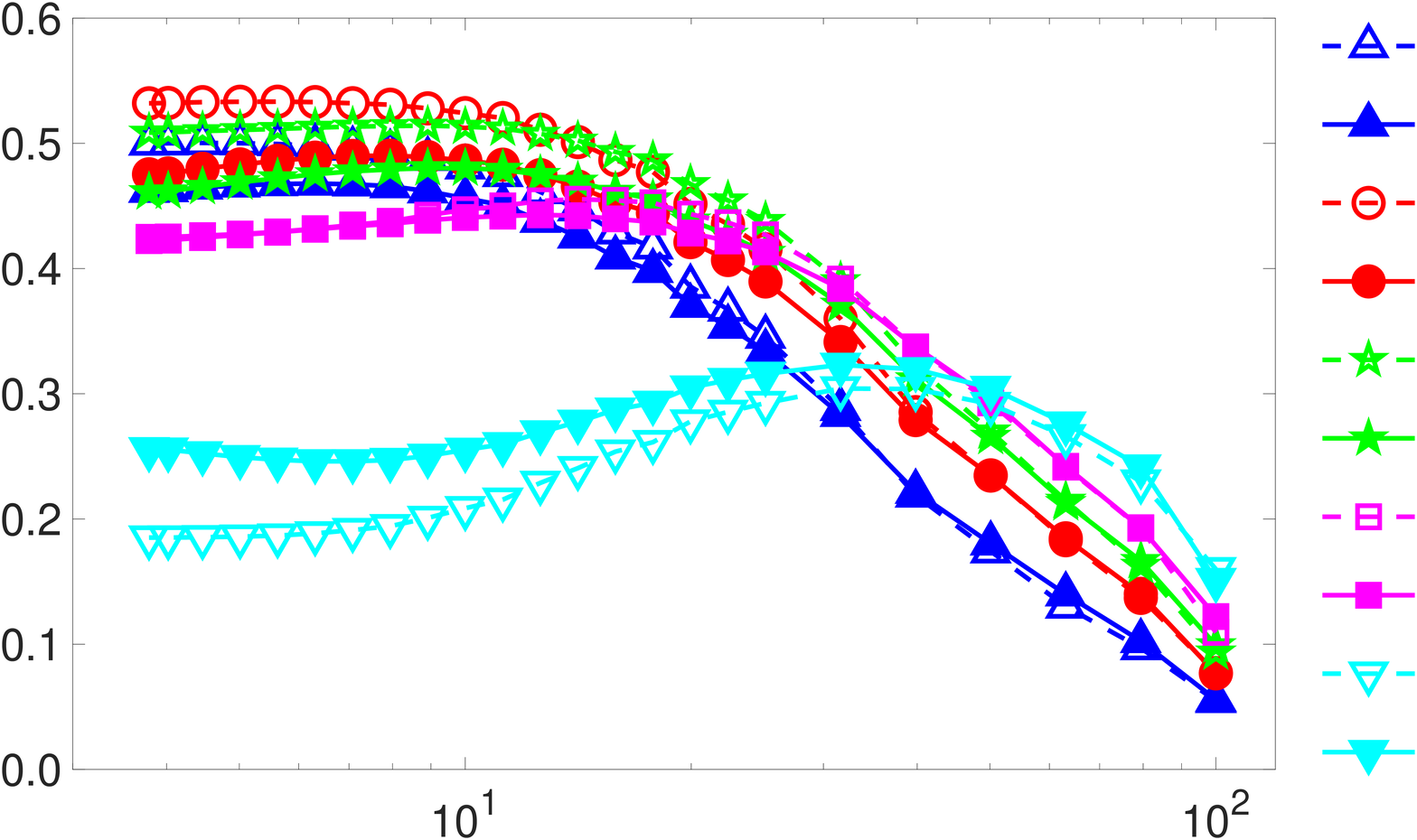}
\put(45,1){\large{$\ell_F / \eta$}}
\put(-8,8){\rotatebox{90}{\large{$\Bigl<\widetilde{\mathcal{Q}}^p(t)\Bigr> \Big/ \sqrt{\langle[\widetilde{\mathcal{Q}}^p(t)]^2\rangle-\langle\widetilde{\mathcal{Q}}^p(t)\rangle^2}$}}}		
\put(87.5,51.85){\large{$St = 0.3$ 1WC}}
\put(87.5,47.05){\large{$St = 0.3$ 2WC}}
\put(87.5,42.25){\large{$St = 0.5$ 1WC}}
\put(87.5,37.45){\large{$St = 0.5$ 2WC}}
\put(87.5,32.65){\large{$St = 0.7$ 1WC}}
\put(87.5,27.85){\large{$St = 0.7$ 2WC}}
\put(87.5,23.05){\large{$St = 1.0$ 1WC} }
\put(87.5,18.25){\large{$St = 1.0$ 2WC}}
\put(87.5,13.45){\large{$St = 2.0$ 1WC}}
\put(87.5,8.65){\large{$St = 2.0$ 2WC}}
\put(51.5,38.5){
\begin{overpic}[trim = 0mm 0mm 0mm 0mm, scale=0.058,clip,tics=20]
{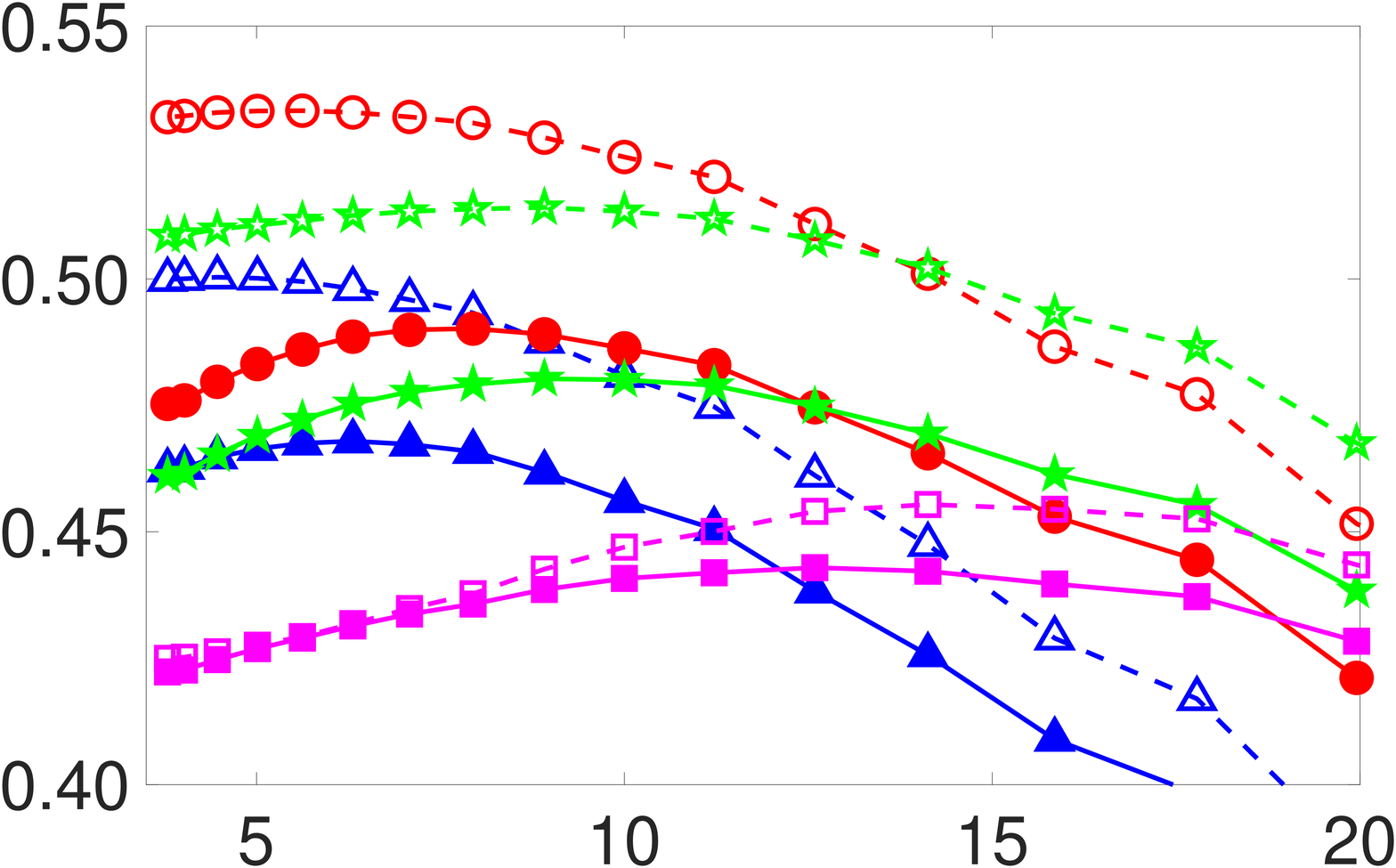}
\end{overpic}
}
	\end{overpic}}
\caption{Ratio of normalized $\langle\widetilde{\mathcal{Q}}^p(t)\rangle$ for the 1WC to 2WC case shown as a function of the normalized filtering length $\ell_F/\eta$ for varying $St$ and fixed $Re_{\lambda} = 87$ and $Fr = 1$.}
\label{Filt_Qnorm_ScaleDep_StDep}
\end{figure}

\begin{figure}
\centering
\vspace{0mm}			
	{\begin{overpic}
	[trim = 0mm -25mm -92mm -3mm,
	scale=0.19,clip,tics=20]{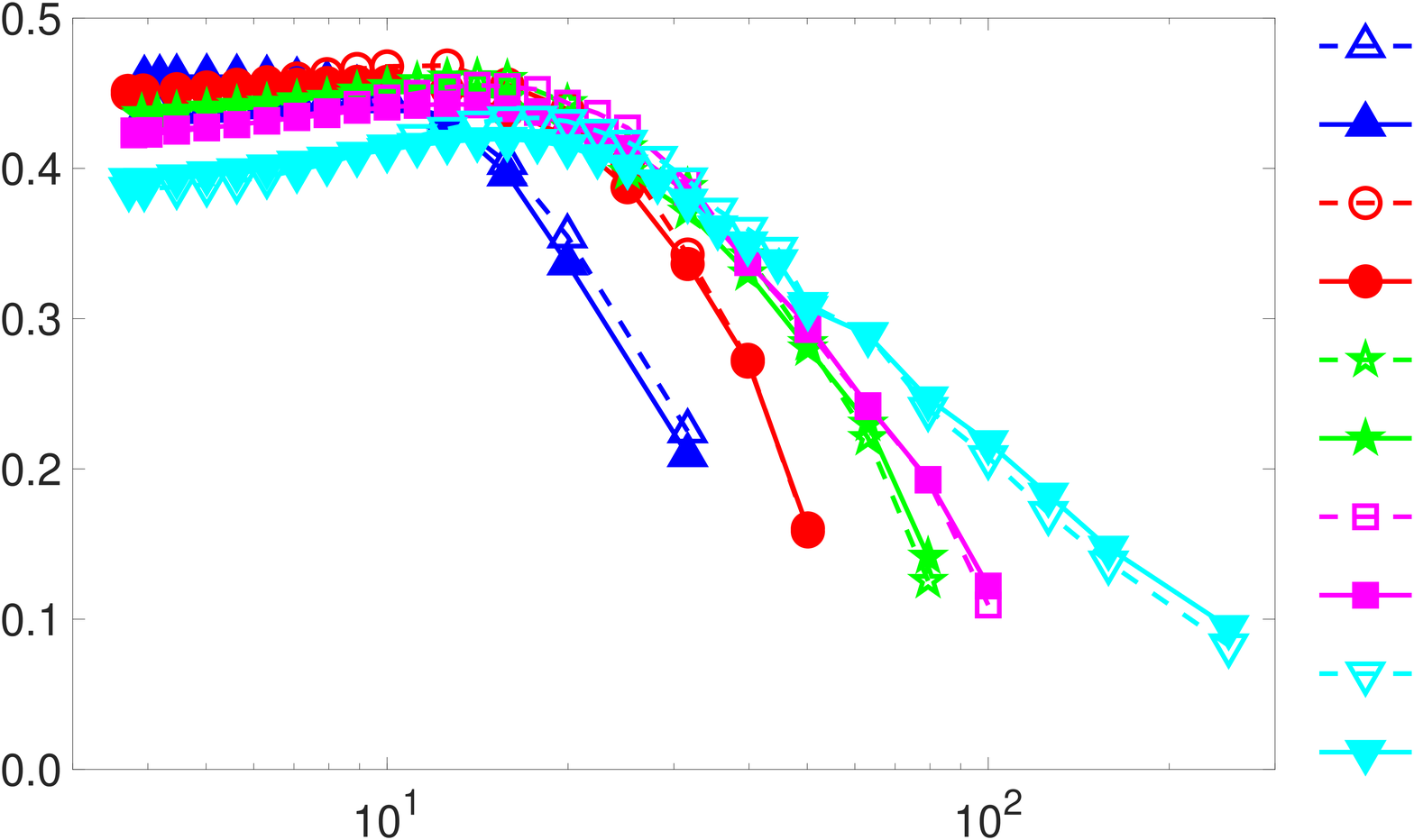}
\put(48,0.5){\large{$\ell_F / \eta$}}
\put(-8,8){\rotatebox{90}{\large{$\Bigl<\widetilde{\mathcal{Q}}^p(t)\Bigr> \Big/ \sqrt{\langle[\widetilde{\mathcal{Q}}^p(t)]^2\rangle-\langle\widetilde{\mathcal{Q}}^p(t)\rangle^2}$}}}			
\put(86.5,51.0){\large{$Re_{\lambda} = 31$ 1WC}}
\put(86.5,46.25){\large{$Re_{\lambda} = 31$ 2WC}}
\put(86.5,41.5){\large{$Re_{\lambda} = 54$ 1WC}}
\put(86.5,36.75){\large{$Re_{\lambda} = 54$ 2WC}}
\put(86.5,32.0){\large{$Re_{\lambda} = 75$ 1WC}}
\put(86.5,27.25){\large{$Re_{\lambda} = 75$ 2WC}}
\put(86.5,22.5){\large{$Re_{\lambda} = 87$ 1WC}}
\put(86.5,17.75){\large{$Re_{\lambda} = 87$ 2WC}}
\put(86.5,13.0){\large{$Re_{\lambda} = 142$ 1WC}}
\put(86.5,8.25){\large{$Re_{\lambda} = 142$ 2WC}}
\put(5.5,10){
\begin{overpic}[trim = 0mm 0mm 0mm 0mm, scale=0.075,clip,tics=20]
{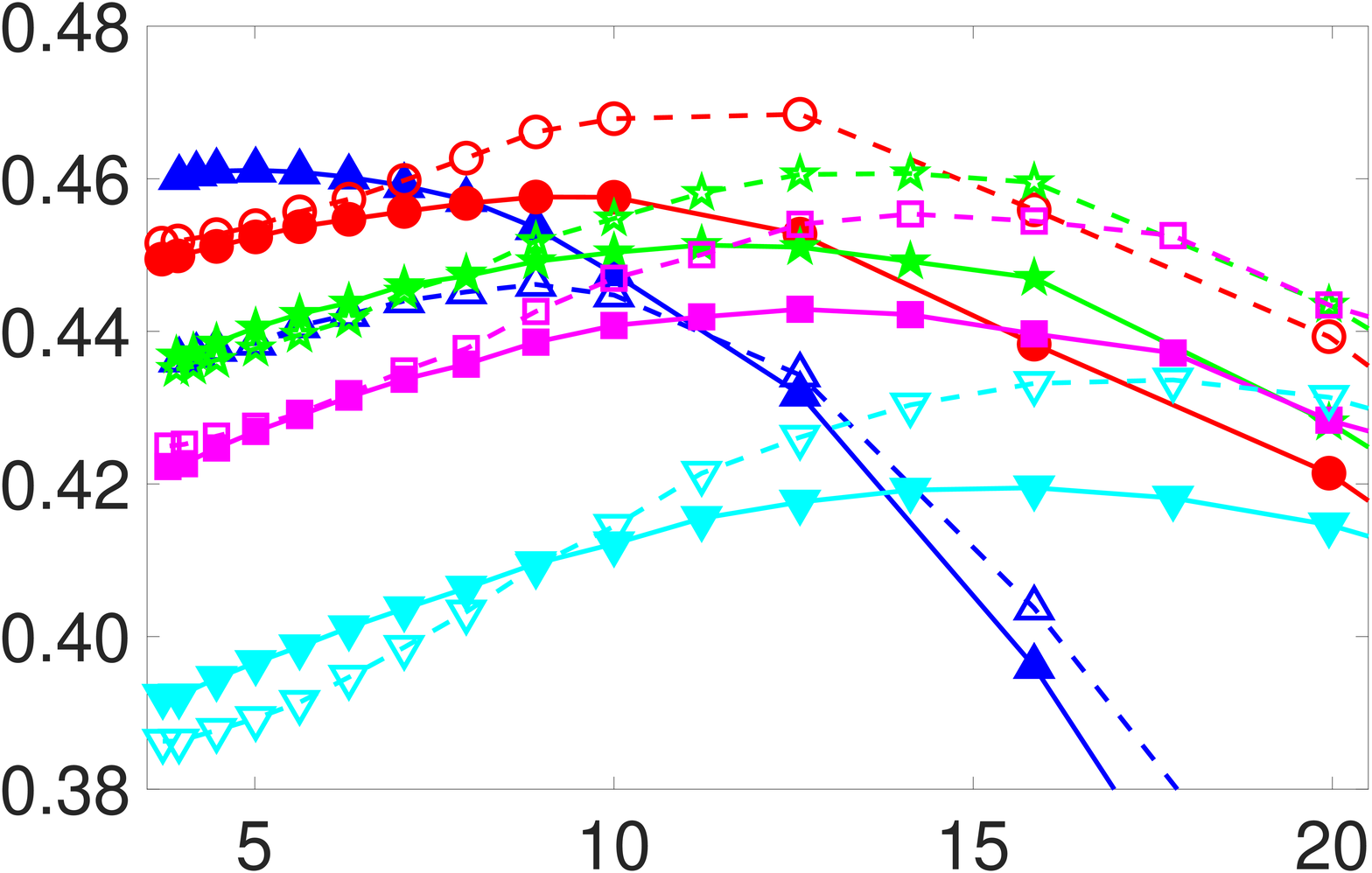}
\end{overpic}
}
\end{overpic}}
\caption{Ratio of normalized $\langle\widetilde{\mathcal{Q}}^p(t)\rangle$ for the 1WC to 2WC case shown as a function of the normalized filtering length $\ell_F/\eta$ for varying $Re_{\lambda}$ and fixed $St = 1$and $Fr = 1$.}
\label{Filt_Qnorm_ScaleDep_ReDep}
\end{figure}
\FloatBarrier

The preferential sweeping mechanism is based in part on the argument that inertial particles preferentially sample strain dominated regions of the flow due to the centrifuge effect \citep{maxey87}. To explore this preferential sampling of the flow in more detail, and its behavior at different scales of the flow, we now introduce the invariant based on the filtered velocity gradients
\begin{align}
\widetilde{\mathcal{Q}}^p(t)\equiv \widetilde{\mathcal{S}}^2(\bm{x}^p(t),t)-\widetilde{\mathcal{R}}^2(\bm{x}^p(t),t),
\end{align}
where $\widetilde{\mathcal{S}}^2\equiv \widetilde{\bm{\mathcal{S}}}\bm{:}\widetilde{\bm{\mathcal{S}}}$, $\widetilde{\mathcal{R}}^2\equiv \widetilde{\bm{\mathcal{R}}}\bm{:}\widetilde{\bm{\mathcal{R}}}$, and $\widetilde{\bm{\mathcal{S}}}$, $\widetilde{\bm{\mathcal{R}}}$ denote $\bm{\mathcal{S}}$, $\bm{\mathcal{R}}$ coarse-grained on the scale $\ell_F$. 
To compare the strength of the preferential sampling of the flow across the scales we consider \[
\Bigl<\widetilde{\mathcal{Q}}^p(t)\Bigr> \Big/ \sqrt{\langle[\widetilde{\mathcal{Q}}^p(t)]^2\rangle-\langle\widetilde{\mathcal{Q}}^p(t)\rangle^2}.\]For homogeneous turbulence, $\langle\widetilde{\mathcal{Q}}^p(t)\rangle=0$ when measured along the trajectories of particles that do not preferentially sample the flow. The normalization involving $\langle[\widetilde{\mathcal{Q}}^p(t)]^2\rangle-\langle\widetilde{\mathcal{Q}}^p(t)\rangle^2$ is used to factor out the trivial scale dependence that can arise simply because the variance of the fluid velocity gradients decreasing with increasing filter scale.

In figure \ref{Filt_Qnorm_ScaleDep_StDep}, we show the normalized $\langle\widetilde{\mathcal{Q}}^p(t)\rangle$ across a range of scales for varying $St$ and fixed $Re_{\lambda}$. The results show $\langle\widetilde{\mathcal{Q}}^p(t)\rangle>0$, indicating that the particles preferentially sample strain dominated regions of the flow at all scales. For larger values of $\ell_F$ the 1WC and 2WC cases are similar, showing that 2WC has a negligible effect on preferential sampling at these scales. When we consider the effect of 2WC when the effect of all scales is included, i.e.~when $\ell_F/\eta \to 0$, then there is a noticeable effect, but the effect is still small. In this case, for fixed $\ell_F$, 2WC slightly reduces $\langle\widetilde{\mathcal{Q}}^p(t)\rangle$ for $St < 1$ and increases $\langle\widetilde{\mathcal{Q}}^p(t)\rangle$ for $St > 1$. For $\ell_F/\eta\lesssim 10$, $\langle\widetilde{\mathcal{Q}}^p(t)\rangle$ is approximately independent of $\ell_F$ when $St < O(1)$, and then decreases for larger $\ell_F$. This can be explained by considering the dependence of preferential sampling on $St_\ell$ and noting that we would expect the preferential sampling at any scale $\ell$ to be maximum for $St_\ell=O(1)$ \citep{tom19}. For $\ell\leq O(10\eta)$, the eddy turnover time $\tau_\ell$ is approximately constant, and hence so also is $St_\ell$. For larger $\ell$, $\tau_\ell$ decreases with increasing $\ell$, so that $St_\ell$ decreases. As such, if $St< O(1)$, then $\langle\widetilde{\mathcal{Q}}^p(t)\rangle$ should be approximately constant for $\ell\leq O(10\eta)$ and will then decrease as $\ell$ is further increased. On the other hand, if $St> O(1)$, then as $\ell$ is increased, $St_\ell$ will decrease and approach $O(1)$, and then continue to reduce a $\ell$ is further increased. In this case, $\langle\widetilde{\mathcal{Q}}^p(t)\rangle$ would have a non-monotonic behavior, first increasing as $\ell$ is increased, and then decreasing, a behavior that can be observed especially for the $St=2$ data in figure \ref{Filt_Qnorm_ScaleDep_StDep}.

The results in figure \ref{Filt_Qnorm_ScaleDep_ReDep} for varying $Re_{\lambda}$ and fixed $St$ also show that 2WC has only a small effect on preferential sampling at all scales. For $\ell_F=0$ (unfiltered results), the preferential sampling becomes weaker as $Re_\lambda$ is increased, consistent with the results of \cite{ireland16a,ireland16b} for $Fr=\infty$. For larger $\ell_F$, the main difference in the results is simply due to the different range of scales available for the particles to preferentially sample as $Re_\lambda$ is varied.

Referring back to the discussion surrounding equation \eqref{Alt} in \S\ref{Theory_2WC_Appl}, the results for $\mathcal{Q}^p(t)$ imply that the principal way that 2WC modifies $\langle u_z(\bm{x}^p(t),t)\rangle$ is by modifying the field $u'_z(\bm{x},t)$ in the vicinity of the particles, and not by modifying the spatial distribution/preferential concentration of the particles associated with the factor $\exp( -\int_0^t\bm{\nabla\cdot}\bm{\mathcal{V}}'(\bm{\mathcal{X}}(s\vert\bm{x},t),s)\,ds)$ in \eqref{Alt}.


%
\section{Conclusions}\label{sec:Conclusions}
In this paper, we have explored how two-way coupling (2WC) affects the preferential sweeping mechanism which is responsible for the enhanced settling speeds of inertial particles in one-way coupled (1WC) turbulent flows. Most numerical and theoretical studies on particle settling have focused on the 1WC regime, where the forcing from the particles on the flow is negligible. \citet{maxey87} developed a theoretical analysis to explain enhanced particle settling in 1WC homogeneous turbulence, and according to this analysis, inertial particles tend to be swept around vortices in the flow due to their inertia, and in the presence of gravity, they exhibit a tendency to be swept around the downward moving side of vortices. This tendency to preferentially sample strain dominated regions of the flow where fluid velocity is pointing in the direction of gravity then generates enhanced settling velocities in turbulence, and is referred to as the preferential sweeping mechanism \citep{wang93}. However, the theoretical analysis of \citet{maxey87} was restricted to $St \ll 1$. In a recent work, \citet{tom19} extended the analysis to arbitrary $St$ which revealed the truly multiscale nature of the preferential sweeping mechanism. The analysis showed that only a restricted range of scales contribute to enhanced settling, and this range depends on particle inertia. The results also showed that the largest scale that contributes to the sweeping is much larger than previously thought. In this work, we applied our analysis to the 2WC case and using DNS demonstrated that preferential sweeping is still the mechanism generating enhanced particle settling in 2WC homogeneous turbulent flows.  

Our DNS considered a volume fraction of $\Phi_v=1.5\times 10^{-5}$ in order to explore the regime where the effect of 2WC on the global fluid statistics is weak, but where 2WC may nevertheless be important for the particle settling. In agreement with previous results for a similar region \citep{bosse06,monchaux17,rosa21}, our results show that 2WC can lead to particle settling enhancements that are more than twice those in the corresponding 1WC case, even though $\Phi_v$ is small. This is an important point since the effects of 2WC are traditionally ignored in weather forecasting models that parameterize particle settling through the atmosphere \citep{kukkonen12} upon the assumption that $\Phi_v$ is sufficiently small in those contexts. As emphasized in \cite{monchaux17}, the key point is that even if $\Phi_v$ is small enough for the particles to not affect the global fluid statistics, it may be strong enough to significantly modify the flow in the vicinity of the particles, which in turn modifies the particle motion. Moreover, recent field experiments \citep{li21a} suggest that turbulence is a key factor influencing the fall speed of snow through the atmosphere, which in turn influences local hydrology and vegetation development \citep{lehning08}. Hence, a detailed understanding of the physical mechanisms responsible for settling enhancement and the impact of 2WC is important not only from a fundamental physics perspective, but also for developing accurate parametrizations of particle settling in atmospheric and weather forecasting models.  

Next, we explored in detail how 2WC modifies the contribution from different scales of the flow to the turbulence-induced enhancement of the particle settling velocities by considering the contribution of scales of size $<\ell_F$ (i.e. sub-grid velocity field) to the particle settling velocity. The results showed that at the smallest scales of the flow, 2WC very strongly enhances the contribution of these scales to the particle settling velocity, with the impact of 2WC decreasing as $\ell_F$ is increased, but still remaining significant for $\ell_F$ approaching the largest scales of the flow. By also considering the second moment of the sub-grid fluid velocity at the particle positions for the 1WC and 2WC cases, we concluded that this strong enhancement due to 2WC is plausibly explained in terms of the fluid dragging effect described by \cite{monchaux17}. According to this effect, the settling inertial particles on average accelerate the flow in their vicinity in the direction of gravity, which causes the fluid velocity in their vicinity to be increased compared with the 1WC case.

To understand the role being played by the preferential sweeping mechanism in the 2WC case, and whether its effect is being suppressed or even eradicated due to the fluid dragging effect, as suggested by \cite{monchaux17}, we considered the average fluid velocity at the particle position conditioned on the local flow structure. Consistent with the preferential sweeping mechanism, the results showed that the average fluid velocity at the particle position (which is the term that quantifies the enhanced particle settling due to turbulence) is dominated by contributions where the particles are located in strain dominated regions of the flow. This is however enhanced in the 2WC case due to the fluid dragging effect. Therefore, rather than 2WC eliminating the importance of preferential sweeping, what actually occurs is more subtle. In particular, for both 1WC and 2WC flows, the settling enhancement due to turbulence is dominated by contributions from particles in strain dominated regions of the flow, but for the 2WC case, in these strain dominated regions the particles also drag the fluid with them, increasing the local fluid velocity and hence leading to further enhancement of the particle settling due to turbulence. Note that the particles also drag the fluid down with them when they are in rotation dominated regions, but again, the contribution from particles in strain dominated regions dominates the settling behavior, consistent with the preferential sweeping mechanism.

We present an alternate interpretation for the results in \citet{monchaux17} regarding the effectiveness of the preferential sweeping mechanism in 2WC flows. As shown in \cite{tom19}, when the particle settling number $Sv$ increases, the particles begin to preferentially sample the small scales less and larger scales more since settling reduces the eddy turnover timescales seen by the particles. For sufficiently large $Sv$, even though particles do not preferentially sample the unfiltered fluid velocity gradient field, they do preferentially sample
the fluid velocity gradient field at some finite filter length $\ell_F$. Hence, the fact that \citet{monchaux17} did not observe preferential sampling of the unfiltered velocity gradients does not imply that the preferential sweeping was ineffective in their 2WC DNS. To observe the preferential sampling in their flow they would have needed to consider the statistics of the filtered fluid velocity gradient along the inertial particle trajectory. Further supporting this argument is that our results for the probability density function and average of the second invariant of fluid velocity gradients sampled by the inertial particles reveals a weak influence of 2WC on these quantities. 

In summary, our results reveal that for the $\Phi_v$ considered, 2WC enhances the particle settling velocities compared to the 1WC case by strongly modifying the fluid velocity in the vicinity of the particles, while only weakly affecting the preferential sampling of the velocity gradients by the inertial particles. As such, these results reveal that the multiscale preferential sweeping continues to be the mechanism by which turbulence enhances the settling of inertial particles in 2WC homogeneous turbulent flows. Important issues for future investigation are how these conclusions might change for larger particle mass loadings, and for smaller Froude numbers.


\backsection[Funding]
{
This work was supported by the National Aeronautics and Space Administration Weather and Atmospheric Dynamics program (grant number NASA 80NSSC20K0912). The computational resources used were provided by the Extreme Science and Engineering Discovery Environment (XSEDE) under allocation CTS170009, which is supported by National Science Foundation (NSF) grant number ACI-1548562 \citep{xsede}. Specifically, the Expanse cluster operated by San Diego Supercomputer Center (SDSC) was used to obtain the results in this work. The Duke Computing Cluster (DCC) operated by Duke University Research Computing was also used to obtain some of the preliminary results for this study.}

\backsection[Declaration of interests]
{ 
The authors report no conflict of interest.
}

\backsection[Author ORCID]{Josin Tom, https://orcid.org/0000-0002-2717-089X, Maurizio Carbone, https://orcid.org/0000-0003-0409-6946, Andrew D. Bragg, https://orcid.org/0000-0001-7068-8048.}


\bibliographystyle{jfm}
\bibliography{jfm}


\end{document}